\documentclass{article}
\usepackage{amssymb}
\usepackage{amsfonts}
\usepackage{amsmath}
\usepackage[dvips]{graphicx}

\begin{document}

\title{Scattering of the acoustic waves on finite cylindrical covers}
\author{Demchuk V.I.*, Datsko B.J.*, Ponomaryov A.N.* \\
*{\small Institute for Applied Problems in Mechanics and Mathematics,}\\
{\small National Academy of Sciences of Ukraine, Lviv, Ukraine}}
\maketitle

\bigskip The interaction problems of wave fields with resilient bounded
bodies arise in the different fields of mechanics, hydro acoustics,
geophysics and seismology.

The analytical solutions of the problems of hydro acoustic waves scattering
by resilient bodies are directly connected with the methods development of
modeling wave fields interaction processes with underwater objects.
Increasing the complexity of underwater objects geometry we get the increase
of mathematical complexity of the pressure field determination problem
solution. In this connection the problems are solved with simplified
mathematical statement, which analysis helps to indicate the significant
features of geometry and structure of modeling object. Up till now number of
analytical and experimental researches were implemented according to the
statement of the problem. These researches were directed to the conformities
determination of influence of any construction or technologic property on
structure parameters of reflected signals.

The problems of scattering of acoustic pressure waves by resilient bodies
with complex geometry have no analytical solution.

Applying the direct numerical methods (e.g. finite difference method) to
solve partial differential equations of these problems is not expedient even
using powerful computers due to multi-dimensional type of the problem. For
this reason high-frequency and low-frequency methods are often used to solve
such problems. But when wavelength is comparable by order with the
characteristic size of disperse body the frequency lies in moderate range.
Exactly this range as shown in \cite{24} is highly informational.

To solve the scattering problems in the moderate frequency range the limited
integral equations method is applied.

Applying Grin theorem let us reduce the solution of stationary problem of
resilient body dispersion to the solution of combined singular equations for
the system of potential displacements. The solution of the above determines
the surface potentials, which are used to define the outside field. Thus,
the number of dimension variables is reduced and it becomes possible to use
numerical methods.

In works \cite{26, 2} the problem of sound diffraction on the finite
cylinder with mixed boundary conditions is solved by Grin's function method.
In \cite{27} the axis-symmetrical dispersed pressure of resilient disk is
computed using finite element method and HelmHoltz integral equation.

As is well known, there is a tight relation between the solution of
scattering and emission problems. It could be defined by reciprocity theorem
\cite{28}. This relation could be used to transfer emission problem results
to scattering problems and vise versa \cite{29}.

While solving complex problems, for example three-dimensional problems of
acoustic waves scattering on the resilient bodies or two-dimensional
problems of scattering on the piece-wise-smooth resilient bodies and problems
of scattering of acoustic waves on the thin-wall covers, troubles with
definite satisfaction of the appropriate boundary conditions arise.

The considered work is devoted to the research of scattered and emissed
acoustic waves by finite cylindrical covers, resilient cylinders and round
plates.

The presence of ribs fracture line in bottoms interface and cylindrical
covers on the face of resilient cylinder and on edge of round plates inputs
the additional troubles of evaluating of such problems.

During last years for the solution of space problems of scattering of
acoustic waves numerical and integral equations methods are effectively
applied. Despite the achieved researches of applying these methods to
three-dimensional problems, questions of efficiency increase of existing
methods and their PC-realization are still actual. In particular, one of the
hydro-acoustic problems is the evaluation of scattered and emitted pressure
field by three-dimensional bodies of cylindrical form. But during the
scattered pressure field exploration near bounded disperser with edges and
also internal features of dissipater troubles arise, which are hardly solved
without analytical research. That is why the important problem of
analytic-numerical methods development and development of software for
acoustic space interaction with finite bodies of cylindrical form are actual.

During the solution of problems of emission and scattering of acoustic waves
by bounded bodies of cylindrical form we use the strategy based on applying
of bounded integral equations method combined with series method and later
using the improvement of convergence method, which consider particularities
of desired functions. The proposed strategy is authorized by the fact that
the series method is effective to solution of equations of resiliency
theory, thin covers and plates, and the bounded integral equations method is
effective to research of infinite acoustic environment vibrations. Using
this methodology let us describe the behavior of bounded resilient bodies
near salient line, explore the pressure field in near and far zone, obtain
the solution of the problem with defined precision in wide frequency range.

In the first chapter basic linear equations of motion and initial relations
are considered, statement and solution of problem of scattering of flat
acoustic wave by finite cylindrical cover bounded at the ends by resilient
bottoms is done.

The improvement of convergence of obtained infinite systems of linear
algebraic equations was held, using asymptotic properties of Fourier
decomposition coefficients of the desired functions while solving the
equations of cover theory.

The solutions of the problem of scattering of flat pressure wave by open-end
free cylindrical cover, and also by tension cylindrical cover with defined
inner tensions. In case of thin resilient cylindrical cover and round plates
its motion is modeled by the linear Thimoshenko model.

Today the problem of learning of dynamical processes of acoustic fields
interaction with thin-walled structures in liquids is of great interest. The
theoretical exploration of these processes is strongly associated with
building of effective analytic-numerical methods of mathematical physics
equation solving, which are realized on today's PC. For example, while
analyzing wave diffraction on solid bodies and covers of revolution with
fixed ends in \cite{33} finite-element and spline-functions methods based on
function approximation theory were applied.

We assume that scattering objects could easily move in boundless liquid
environment or to be deformless.

In this work we propose the analytical method of dynamical problems solution
of fixed oscillation of infinite acoustic environment interacted with
resilient finite cylindrical bodies.

In the nowadays technics problems of oscillation of finite cylindrical
covers, filled with liquid, their sound waves emission at vibration in
infinite acoustic environment are actual.

While interaction of wave fields with resilient bounded bodies problems
arise in different fields of technics, medicine, hydro-acoustics and
geo-physics.

The analytical solutions of problems of \ scattering of acoustic waves by
resilient bodies are directly related with developing methods of modelling
of interaction processes of wave fields with underwater objects.

\begin{figure}[htbp]
\begin{center}
\includegraphics[bb=10 397 394 829,width=3in]{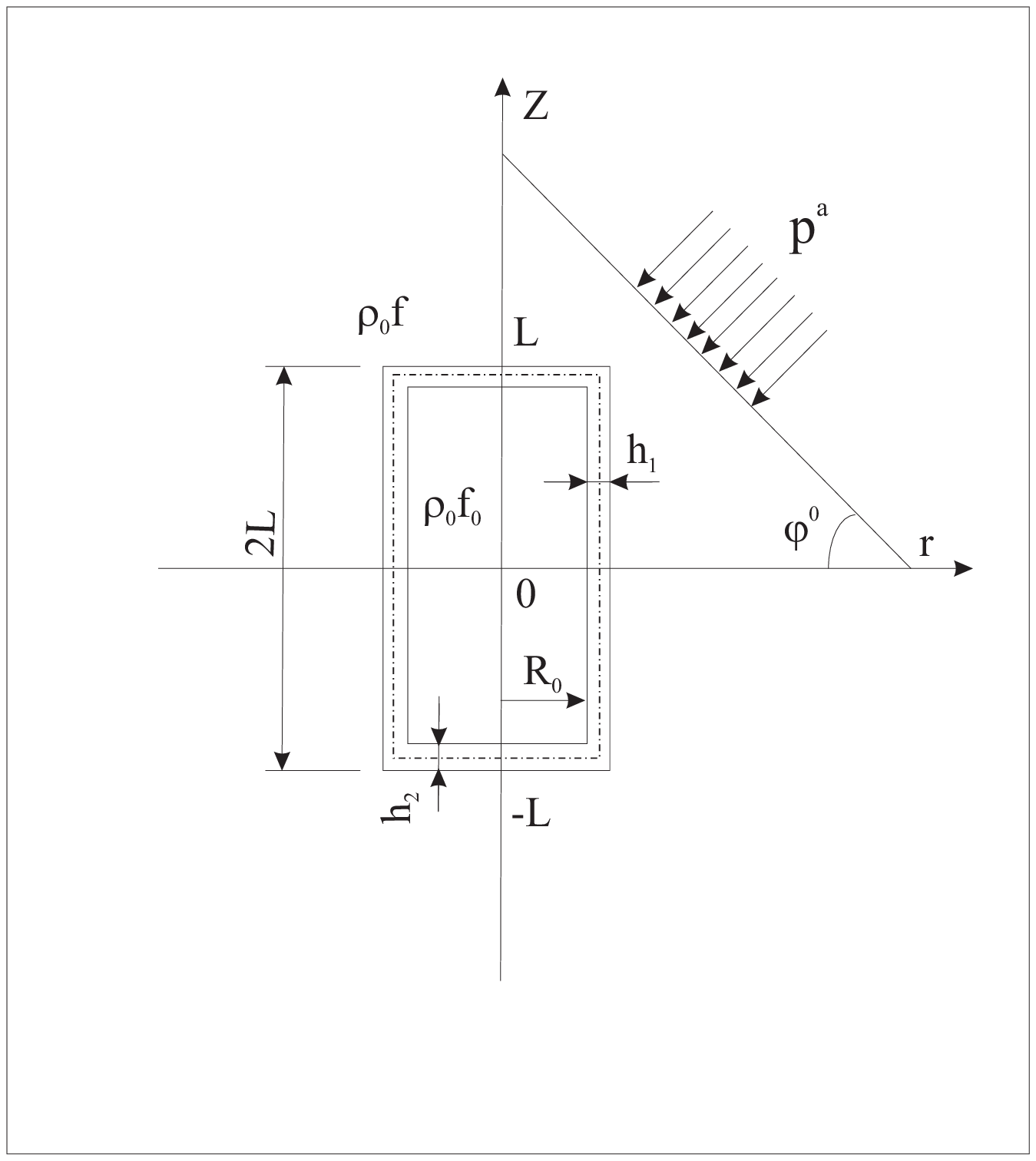} 
\end{center}
\par
\centerline{Fig. 1}
\end{figure}


Let in the infinite ideal compressible rigid media the cylindrical cover of
finite length bounded at the ends by bottoms is placed. Cover is filled with
ideal compressible fluid inside with density $\rho _{0}$ and sound velocity $%
c_{0}$.

The considered cover is referred to cylindrical system of coordinates $%
r^{\prime },z^{\prime },\Theta $ where axis $oz^{\prime }$ is aligned with
axis of cylindrical cover and plane $z^{\prime }=0$ is equidistant to ends
(see Fig. 1). Flat harmonic in time pressure wave falls to the cover (see
Fig. 1):

\begin{equation}
P^{\alpha }\left( r,z,\Theta ,\tau \right) =P_{0}\exp \left( -i\omega \left(
r\cos \Theta \sin \varphi ^{\ast }+z\cos \varphi ^{\ast }-\tau \right)
\right)  \label{1}
\end{equation}%
In formula (\ref{1}) and below following symbolism is used: $r=\frac{%
r^{\prime }}{R_{0}};z=\frac{z^{\prime }}{R_{0}}$ are nondimentional
coordinates, $\tau =\frac{ct}{R_{0}};\omega =\frac{\Omega R_{0}}{c}$ are
nondimentional time and angular frequency, $c$ is sound velocity in the
dissipating body environment, $R_{0}$ is radius of middle surface of
cylindrical cover, $t,\Omega $ are time and angular frequency, $\varphi
^{\ast }$ is angle between wave direction and axis of the cylindrical cover,
$P_{0}$ is constant with the dimension of pressure. Time factor $\exp \left(
i\omega \tau \right) $ is omitted below.

The problem of definition of dissipated pressure field outside the cover $%
P^{e}\left( r,z,\Theta \right) $ and in filler $P^{0}\left( r,z,\Theta
\right) $ and components of cover movement vector $\overset{\rightarrow }{U}$
and plates $\overset{\rightarrow }{W}$ is reduced to solution of system of
differential equations.

Dissipated acoustic pressure $P^{e}\left( r,z,\Theta \right) $ in unbounded
acoustic medium is described by wave equation \cite{2}%
\begin{equation}
\left( \Delta +\varpi ^{2}\right) P^{e}\left( r,z,\theta \right) =0
\label{2}
\end{equation}%
where $\Delta $ is three-dimensional Laplace operator.

Dissipated pressure in fluid which fills the cover $P^{0}\left( r,z,\Theta
\right) $ satisfies the wave equation \cite{3}%
\begin{equation}
\left( \Delta +\frac{\varpi ^{2}}{\beta _{0}^{2}}\right) P^{o}\left(
r,z,\theta \right) =0  \label{3}
\end{equation}%
where $\beta _{0}^{2}=\frac{c^{2}}{c_{0}^{2}}.$

The dynamics of thin resilient cylindrical cover is described by linear
equation of Timoshenko cover theory which takes into account rotary inertia
and deformation of transversed shift \cite{3}
\begin{equation}
L_{ij}u_{j}=g_{1}\delta _{i3};\left( L_{ij}=L_{ji};i,j=1,2,3,4,5\right)
\label{4}
\end{equation}%
where $\delta _{ij}$ is the Cronicler's symbol,
\begin{eqnarray*}
L_{11} &=&\left( 1+a^{2}\right) \left( \frac{\partial ^{2}}{\partial \theta
^{2}}-\varkappa _{1}^{2}\right) +\sigma _{1}\frac{\partial ^{2}}{\partial
z^{2}}+\frac{\varpi ^{2}}{\beta ^{2}}; \\
L_{12} &=&a^{2}\left( -\frac{\partial ^{2}}{\partial \theta ^{2}}+\varkappa
_{1}^{2}+\sigma _{1}\frac{\partial ^{2}}{\partial z^{2}}+\frac{\varpi ^{2}}{%
\beta ^{2}}\right) +\varkappa _{1}^{2}
\end{eqnarray*}%
\begin{eqnarray*}
L_{22} &=&a^{2}\left( \frac{\partial ^{2}}{\partial \theta ^{2}}-\varkappa
_{1}^{2}+\sigma _{1}\frac{\partial ^{2}}{\partial z^{2}}+\frac{\varpi ^{2}}{%
\beta ^{2}}\right) -\varkappa _{1}^{2}; \\
L_{13} &=&\left( 1+a^{2}\right) \left( 1+\varkappa _{1}^{2}\right) \frac{%
\partial }{\partial \theta };L_{14}=\sigma _{2}\frac{\partial ^{2}}{\partial
z\partial \theta };L_{15}=L_{24}=0;
\end{eqnarray*}%
\begin{eqnarray*}
L_{23} &=&\left[ \left( 1+a^{2}\right) \varkappa _{1}^{2}+a^{2}\right] \frac{%
\partial }{\partial \theta };L_{25}=a^{2}\sigma _{2}\frac{\partial ^{2}}{%
\partial z\partial \theta }; \\
L_{33} &=&\left( 1+a^{2}\right) \left( -\varkappa _{1}^{2}\frac{\partial ^{2}%
}{\partial \theta ^{2}}+1\right) -\varkappa _{1}^{2}\frac{\partial ^{2}}{%
\partial z^{2}}-\frac{\varpi ^{2}}{\beta ^{2}};L_{34}=\nu _{01}\frac{%
\partial }{\partial z};
\end{eqnarray*}%
\begin{eqnarray*}
L_{35} &=&-\varkappa _{1}^{2}\frac{\partial }{\partial z};L_{44}=\left(
1+a^{2}\right) \sigma _{1}\frac{\partial ^{2}}{\partial \theta ^{2}}+\frac{%
\partial ^{2}}{\partial z^{2}}+\frac{\varpi ^{2}}{\beta ^{2}}; \\
L_{45} &=&a^{2}\left( -\sigma _{1}\frac{\partial ^{2}}{\partial \theta ^{2}}+%
\frac{\partial ^{2}}{\partial z^{2}}+\frac{\varpi ^{2}}{\beta ^{2}}\right)
;\sigma _{1}=\frac{1-\nu _{01}}{2};\sigma _{2}=\frac{1+\nu _{01}}{2};
\end{eqnarray*}%
\begin{eqnarray*}
L_{55} &=&a^{2}\left( \sigma _{1}\frac{\partial ^{2}}{\partial \theta ^{2}}+%
\frac{\partial ^{2}}{\partial z^{2}}+\frac{\varpi ^{2}}{\beta ^{2}}\right)
;\beta ^{2}=\frac{c_{10}^{2}}{c^{2}};\varkappa _{1}=\frac{c_{20}}{c_{10}}; \\
g_{1} &=&\frac{g_{r}}{\xi \rho c^{2}};\xi =\frac{\beta ^{2}h_{1}\rho _{1}}{%
R_{0}\rho };
\end{eqnarray*}%
\begin{equation*}
a^{2}=\frac{h_{1}^{2}}{R_{0}^{2}};c_{10}^{2}=\frac{E_{1}}{\rho \left( 1-\nu
_{01}^{2}\right) };c_{20}^{2}=\frac{E_{1}k_{T1}}{2\rho _{1}\left( 1+\nu
_{01}\right) }.
\end{equation*}

$E_{1},\nu _{01},\rho _{1},h_{1}$ is Young's modulus, Poisson's ratio,
density and thickness of cover material, $\rho $ is density of outside
liquid medium, $U_{2},U_{4}$ are turning angle of normal of cover's middle
surface in planes $r\Theta $ and $rz$ appropriately; $g_{r}$ is radial
component of outside force falls on middle surface area unit.

Differential equations of Timoshenko-Mindlin which describe the transverse
motion of round resilient plates with shifted and rotary inertia taken into
account and characterizing few first asymmetrical thickness mod oscillations
have the following form \cite{4}%
\begin{equation}
\left( \Delta _{0}^{2}+\alpha _{1}\Delta _{0}-\alpha _{2}\right) W_{z}\left(
r,\theta \right) =\left( \alpha _{2}^{\ast }-\alpha _{1}^{\ast }\Delta
_{0}\right) g_{z}  \label{5}
\end{equation}%
\begin{equation}
\left( \Delta _{0}+k_{3}^{2}\right) \Phi \left( r,\theta \right) =0
\label{6}
\end{equation}%
where
\begin{eqnarray*}
\Delta _{0} &=&\frac{\partial ^{2}}{\partial r^{2}}+\frac{1}{r}\frac{%
\partial }{\partial r}+\frac{1}{r^{2}}\frac{\partial ^{2}}{\partial \theta
^{2}};\alpha _{1}=\frac{mc^{2}\omega ^{2}\left( S+I\right) }{D_{2}}; \\
\alpha _{2} &=&\frac{mc^{2}\omega ^{2}}{D_{2}}\left( 1-\frac{mc^{2}\omega
^{2}SI}{D_{2}}\right) ;\alpha _{1}^{\ast }=\frac{S}{D_{2}};
\end{eqnarray*}%
\begin{eqnarray*}
m &=&\frac{h_{2}\rho _{2}}{R_{0}};\alpha _{2}^{\ast }=\frac{1}{D_{2}}\left(
1-\frac{mc^{2}\omega ^{2}SI}{D_{2}}\right) ;S=\frac{E_{2}h_{2}^{2}}{%
G_{2}R_{0}^{2}k_{T2}\left( 1-\nu _{02}^{2}\right) }; \\
I &=&\frac{h_{2}^{2}}{12R_{0}^{2}};D_{2}=\frac{E_{2}h_{2}^{3}}{%
12R_{0}^{2}\left( 1-\nu _{02}^{2}\right) };
\end{eqnarray*}%
\begin{equation*}
k_{3}^{2}=\frac{\varpi ^{2}}{\beta _{3}^{2}}-\frac{2}{S\left( 1-\nu
_{02}\right) };\beta _{3}^{2}=\frac{c^{2}}{c_{22}^{2}};c_{22}^{2}=\frac{%
E_{2}k_{T2}}{2\rho _{2}\left( 1+\nu _{02}\right) };
\end{equation*}%
$g_{z}$ - normal component of outside force, which falls falls on middle
surface area unit; $E_{2},\nu _{02},G_{2},\rho _{2},h_{2}$ is Young's
modulus, Poisson's ratio, shift ratio, density and plate thickness
appropriately, $K_{T2}$ is numerical shift coefficient; $\Phi $ is auxiliary
function; $C_{22}$ is velocity of transversed waves in plate.

Let us define the radial and tangent components of motion vector of plates $%
W_{r},W_{\Theta }$ which characterize the flat stress state as follows \cite%
{3, 5}%
\begin{equation}
W_{z}=\frac{\partial \varphi }{\partial r}+\frac{1}{r}\frac{\partial \Psi }{%
\partial r};W_{\theta }=\frac{1}{r}\frac{\partial \varphi }{\partial \theta }%
-\frac{\partial \Psi }{\partial r};  \label{7}
\end{equation}%
Here the scalar potential $\varphi $ and non-zero component of vector
potential $\psi $ satisfy the wave equations
\begin{equation}
\left( \Delta _{0}+\frac{\omega ^{2}}{\beta _{1}^{2}}\right) \varphi \left(
r,\theta \right) =0;\left( \Delta _{0}+\frac{\omega ^{2}}{\beta _{2}^{2}}%
\right) \Psi \left( r,\theta \right) =0  \label{8}
\end{equation}%
where
\begin{equation*}
\beta _{1}=\frac{c_{11}}{c};\beta _{2}=\frac{c_{21}}{c};c_{11}^{2}=\frac{%
E_{2}\left( 1-\nu _{02}\right) }{\rho \left( 1+\nu _{02}\right) \left(
1-2\nu _{02}\right) };c_{21}^{2}=\frac{E_{2}}{2\rho \left( 1-\nu
_{02}\right) }.
\end{equation*}%
The solution of differential equations (\ref{2})-(\ref{8}) must satisfy the
following boundary conditions \cite{3, 7}

a) continuity translation on middle cover surface and plates
\begin{equation}
\left[ \frac{\partial }{\partial r}\left( P^{a}+P^{e}\right) \right]
_{r=1}=\rho \omega ^{2}c^{2}u_{3};g_{r}=\left[ -P^{a}-P^{e}+P^{o}\right]
_{r=1}  \label{9a}
\end{equation}%
\begin{equation*}
\left[ \frac{\partial }{\partial r}P^{o}\right] _{r=1}=\xi _{01}\beta
_{0}^{-1}\rho \omega ^{2}c^{2}u_{3};\xi _{01}=\frac{\beta _{0}^{2}h_{1}\rho
_{1}}{R_{0}\rho };
\end{equation*}%
\begin{equation*}
\left[ \frac{\partial }{\partial z}\left( P^{a}+P^{e}\right) \right]
_{z=l}=\rho \omega ^{2}c^{2}W_{z};g_{z}=\left[ -P^{a}-P^{e}+P^{o}\right]
_{z=_{-}^{+}l};
\end{equation*}%
\begin{equation*}
\left[ \frac{\partial }{\partial z}P^{o}\right] _{z=l}=\xi _{01}\beta
_{0}^{-1}\rho \omega ^{2}c^{2}W_{z};\xi _{02}=\frac{\beta _{0}^{2}h_{2}\rho
_{2}}{R_{0}\rho };
\end{equation*}%
\qquad \qquad\ \ \

b) continuity translation in junction place of cover and plates
\begin{equation}
\left[ u_{3}\right] _{z=l}=\left[ W_{r}\right] _{r=1};\left[ u_{5}\right]
_{z=l}=\left[ W_{z}\right] _{r=1};\left[ u_{1}\right] _{z=l}=\left[
W_{\theta }\right] _{r=1};  \label{9b}
\end{equation}

c) continuity of turning angle $\left( r=1,z=\pm l\right) $%
\begin{equation}
\left[ u_{2}\right] _{z=l}=\left[ \frac{1}{R_{0}}\frac{\partial W_{r}}{%
\partial r}-\frac{W_{\theta }}{R_{0}}\right] _{r=1};\left[ u_{4}\right]
_{z=l}=\left[ -\frac{\partial G^{\ast }}{\partial r}+\frac{1}{r}\frac{%
\partial \Phi }{\partial r}\right] _{r=1}  \label{10}
\end{equation}

d) continuity of forces $\left( r=1,z=l\right) $%
\begin{equation*}
B_{1}\left[ \frac{\partial u_{5}}{\partial z}+\frac{\nu _{01}}{R_{0}}\left(
\frac{\partial u_{1}}{\partial \theta }+u_{3}\right) \right] =\Lambda _{1}^{%
{\acute{}}%
}\left[ \frac{\partial }{\partial r}\left( G^{\ast }-W_{z}\right) +\frac{1}{r%
}\frac{\partial \Phi }{\partial r}\right] ;
\end{equation*}%
\begin{equation}
\Lambda _{1}^{%
{\acute{}}%
}\left( u_{4}+\frac{\partial u_{3}}{\partial z}\right) =\frac{E_{2}h_{2}}{%
\left( 1-\nu _{02}^{2}\right) R_{0}}\left[ \frac{\partial W_{r}}{\partial r}+%
\frac{\nu _{02}}{r}\left( W_{z}+\frac{\partial W_{\theta }}{\partial \theta }%
\right) \right] ;  \label{11}
\end{equation}%
\begin{equation*}
B_{1}\frac{1-\nu _{01}}{2}\left( \frac{1}{R_{0}}\frac{\partial u_{5}}{%
\partial \theta }+\frac{\partial u_{1}}{\partial z}\right) =\frac{E_{2}h_{2}%
}{\left( 1-\nu _{02}\right) R_{0}}\left( \frac{1}{r}\frac{\partial W_{z}}{%
\partial r}+\frac{\partial W_{\theta }}{\partial r}-\frac{W_{\theta }}{r}%
\right) ;
\end{equation*}

e) continuity of moments $\left( r=1,z=l\right) $%
\begin{equation}
D_{1}\left( \frac{\partial u_{4}}{\partial z}+\frac{\nu _{01}}{R_{0}}\frac{%
\partial u_{2}}{\partial \theta }\right) =-D_{2}\left[ \frac{\partial
^{2}G^{\ast }}{\partial r^{2}}+\nu _{02}\left( \frac{1}{r}\frac{\partial
G^{\ast }}{\partial r}+\frac{1}{r^{2}}\frac{\partial ^{2}G^{\ast }}{\partial
r^{2}}\right) -\frac{1-\nu _{02}}{2}\frac{\partial \Phi }{\partial r}\right]
;  \label{12}
\end{equation}%
\begin{equation*}
D_{1}\frac{1-\nu _{01}}{2}\left( \frac{\partial u_{2}}{\partial z}+\frac{1}{%
R_{0}}\frac{\partial u_{4}}{\partial \theta }+\frac{\partial u_{1}}{\partial
z}\right) =
\end{equation*}%
\begin{equation*}
-D_{2}\left( 1-\nu _{02}\right) \left\{ \frac{\partial ^{2}G^{\ast }}{%
\partial r\partial \theta }+\frac{1}{2}\left[ \Delta _{0}\Phi -2\left( \frac{%
1}{r}\frac{\partial \Phi }{\partial r}+\frac{1}{r^{2}}\frac{\partial
^{2}\Phi }{\partial \theta ^{2}}\right) \right] \right\}
\end{equation*}%
where
\begin{equation*}
B_{1}=\frac{E_{1}h_{1}}{\left( 1-\nu _{02}^{2}\right) R_{0}};G^{\ast
}=W_{z}+S\Delta _{0}W_{z}+\frac{S^{2}}{D_{2}}g_{z};
\end{equation*}%
\begin{equation*}
\Lambda _{\sigma }^{%
{\acute{}}%
}=\frac{k_{t\sigma }h_{\sigma }}{R_{0}};D_{\sigma }=\frac{E_{\sigma
}h_{\sigma }^{3}}{12\left( 1-\nu _{0\sigma }^{2}\right) R_{0}^{3}};\sigma
=1,2.
\end{equation*}

Taking into account the symmetry relative to plane $\Theta =0$ we find the
solution of equations (\ref{2})-(\ref{8}) in the Fourier series form:
\begin{equation*}
u_{1}\left( z,\theta \right) =\sum_{n=0}^{\infty }\sum_{\nu =0}^{\infty
}\left( u_{1n\nu }^{+}\cos \beta _{\nu }z+u_{1n\nu }^{-}\sin \beta _{\nu
}z\right) \sin n\theta ;
\end{equation*}%
\begin{equation*}
u_{2}\left( z,\theta \right) =\sum_{n=0}^{\infty }\sum_{\nu =0}^{\infty
}\left( u_{2n\nu }^{+}\cos \beta _{\nu }z+u_{2n\nu }^{-}\sin \beta _{\nu
}z\right) \sin n\theta ;
\end{equation*}%
\begin{equation*}
u_{3}\left( z,\theta \right) =\sum_{n=0}^{\infty }\sum_{\nu =0}^{\infty
}\left( u_{3n\nu }^{+}\cos \beta _{\nu }z+u_{3n\nu }^{-}\sin \beta _{\nu
}z\right) \cos n\theta ;
\end{equation*}%
\begin{equation*}
u_{4}\left( z,\theta \right) =\sum_{n=0}^{\infty }\sum_{\nu =0}^{\infty
}\left( u_{4n\nu }^{+}\sin \beta _{\nu }z+u_{4n\nu }^{-}\cos \beta _{\nu
}z\right) \cos n\theta ;
\end{equation*}%
\begin{equation}
u_{5}\left( z,\theta \right) =\sum_{n=0}^{\infty }\sum_{\nu =0}^{\infty
}\left( u_{5n\nu }^{+}\sin \beta _{\nu }z+u_{5n\nu }^{-}\cos \beta _{\nu
}z\right) \cos n\theta  \label{13}
\end{equation}%
\begin{equation*}
W_{z}\left( r,\theta \right) =\sum_{n=0}^{\infty }W_{z,n}\left( r\right)
\cos n\theta ;W_{r}\left( r,\theta \right) =\sum_{n=0}^{\infty
}W_{r,n}\left( r\right) \cos n\theta ;
\end{equation*}

\begin{equation*}
\varphi \left( r,\theta \right) =\sum_{n=0}^{\infty }\varphi _{n}\left(
r\right) \cos n\theta ;\Psi \left( r,\theta \right) =\sum_{n=0}^{\infty
}\Psi _{n}\left( r\right) \sin n\theta
\end{equation*}%
\begin{equation*}
W_{\theta n}\left( r,\theta \right) =\sum_{n=0}^{\infty }W_{\theta ,n}\left(
r\right) \sin n\theta ;\Phi \left( r,\theta \right) =\sum_{n=0}^{\infty
}\Phi _{n}\left( r\right) \sin n\theta
\end{equation*}%
\begin{equation*}
P^{e}\left( r,z,\theta \right) =\sum_{n=0}^{\infty }P_{n}^{e}\left(
r,z\right) \cos n\theta ;P^{0}\left( r,z,\theta \right) =\sum_{n=0}^{\infty
}P_{n}^{0}\left( r,z\right) \cos n\theta
\end{equation*}%
where
\begin{equation*}
\beta _{\nu }=\frac{\nu \pi }{l};l=\frac{L}{R_{0}};
\end{equation*}%
In relations (\ref{13}) \textquotedblright +\textquotedblright\ is the
symmetric and \textquotedblright -\textquotedblright\ is the non-symmetric
components of general problem solution.

Let us exemplify the solution of wave equation (\ref{2}) which satisfies the
Zommerfield's condition in the form of Helmholtz-Huygens integral \cite{8}%
\begin{equation}
P^{e}\left( r,z,\theta \right) =\int_{\sigma _{0}}\left( P^{e}\mid _{\sigma
_{0}}\frac{\partial G}{\partial n_{0}}-\frac{\partial P^{e}}{\partial n}\mid
_{\sigma _{0}}G\right) d\sigma _{0}  \label{14}
\end{equation}%
where $d\sigma _{0}$ is element of dissipating surface area $\sigma _{0};$

$G=\left( 4\pi R^{\ast }\right) ^{-1}\exp \left( -i\omega R^{\ast }\right) $
is the fundamental solution of Helmholtz equation; $R^{\ast
2}=r^{2}+r_{0}^{2}-2rr_{0}\cos \left( \Theta -\Theta _{0}\right) +\left(
z-z_{0}\right) ^{2}$ is the distance between point of observation $\left(
r,z,\Theta \right) $ and point $\left( r_{0},z_{0},\Theta _{0}\right) $ on
the surface $\sigma _{0};$ $\frac{\partial }{\partial n_{0}}$ is the
derivative with respect to outside normal of surface $\sigma _{0}$ in point $%
\left( r,z,\Theta \right) ;$ $P^{e}\left( r_{0},z_{0},\Theta _{0}\right) $
is the value of dissipated pressure on surface $\sigma _{0}.$

Let us arrange the defined representation of function $G$ in cylindrical
coordinates \cite{8} with divided coordinate $z$%
\begin{equation}
G=\frac{1}{4\pi }\sum_{m=0}^{\infty }\epsilon _{m}\cos \left[ m\left( \theta
-\theta _{0}\right) \right] \int_{0}^{\infty }\lambda J_{m}\left( \lambda
r_{0}\right) J_{m}\left( \lambda r\right) \frac{e^{-\varkappa \mid
z-z_{0}\mid }}{\varkappa }d\lambda  \label{15}
\end{equation}%
where \ae =$\left\{
\begin{array}{c}
\sqrt{\lambda ^{2}-\omega ^{2}},\text{ \ \ if }\lambda >\omega \\
i\sqrt{\omega ^{2}-\lambda ^{2}},\text{ \ if }\omega >\lambda%
\end{array}%
\right. ,$ $J_{m}\left( {}\right) $ is Bessel's function of the first kind; $%
\varepsilon _{0}=1;\varepsilon _{m}=2;m\geq 1.$

Let us assume on the surface of closed cylindrical cover the distributed
pressure $P^{e}\left( r_{0},z_{0},\Theta _{0}\right) $ and its formal
derivatives are presented in the following form:

a) by Fourier's series on the side surface of cylinder
\begin{equation}
P^{e}\left( z_{0},\theta _{0}\right) =\sum_{n=0}^{\infty }\sum_{\nu
=0}^{\infty }\left( f_{n\nu }^{\ast +e}\cos \beta _{\nu }z_{0}+f_{n\nu
}^{-e}\sin \beta _{\nu }z_{0}\right) \cos n\theta  \label{16}
\end{equation}%
\begin{equation*}
\frac{\partial }{\partial r}P^{e}\left( z_{0},\theta _{0}\right)
=\sum_{n=0}^{\infty }\sum_{\nu =0}^{\infty }\left( f_{n\nu }^{\ast +e}\cos
\beta _{\nu }z_{0}+f_{n\nu }^{\ast -e}\sin \beta _{\nu }z_{0}\right) \cos
n\theta
\end{equation*}%
b) by Fourier-Bessel's series on plates ends $z_{0}=\pm l$%
\begin{equation*}
P^{e}\left( z_{0},\theta _{0}\right) =\sum_{n=0}^{\infty }\sum_{\nu
=0}^{\infty }g_{nj}^{e\pm }J_{n}\left( \gamma _{jn}r_{0}\right) \cos n\theta
_{0}
\end{equation*}%
\begin{equation}
\frac{\partial }{\partial z}P^{e}\left( z_{0},\theta _{0}\right)
=\sum_{n=0}^{\infty }\sum_{\nu =0}^{\infty }g_{nj}^{\ast e\pm }J_{n}\left(
\gamma _{jn}r_{0}\right) \cos n\theta _{0}  \label{17}
\end{equation}

where $\gamma _{jn}$ is the zeroes of Bessel's function derivative $\left(
J_{n}^{\prime }\left( \theta _{jn}\right) =0\right) $%
\begin{equation*}
g_{nj\mid z_{0=l}}^{e+}=g_{nj\mid z_{0=-l}}^{e+};g_{nj\mid
z_{0=l}}^{e-}=g_{nj\mid z_{0=-l}}^{e-};g_{nj\mid z_{0=l}}^{\ast
e+}=g_{nj\mid z_{0=-l}}^{\ast e+};g_{nj\mid z_{0=l}}^{\ast e-}=g_{nj\mid
z_{0=-l}}^{\ast e-}
\end{equation*}

Let us write the distributed pressure $P^{e}$ as the sum of symmetrical and
antisymmetric components relative to $z=0$%
\begin{equation*}
P^{e}\left( r,z,\theta \right) =P^{e+}\left( r,z,\theta \right)
+P^{e-}\left( r,z,\theta \right)
\end{equation*}%
Substituting the expansions (\ref{15}), (\ref{16}), (\ref{17}) into (\ref{14}%
) and integrating using (\ref{13}) we get the following expressions for the
distributed pressure to region

a)symmetrical component
\begin{equation*}
P_{n}^{e+}\left( r,z\right) =\frac{\varepsilon _{n}}{2}\sum_{\nu =0}^{\infty
}f_{n\nu }^{e+}\int_{0}^{\infty }\frac{\lambda ^{2}j_{n}^{\shortmid }\left(
\lambda \right) J_{n}\left( \lambda r\right) }{\varkappa ^{2}+\beta _{\nu
}^{2}}\left[ \cos \beta _{\nu }z-\left( -1\right) ^{\nu }ch\left( \varkappa
z\right) e^{-\varkappa l}\right] d\lambda -
\end{equation*}%
\begin{equation}
-\frac{\varepsilon _{n}}{2}\sum_{\nu =0}^{\infty }f_{n\nu }^{\ast
e+}\int_{0}^{\infty }\frac{\lambda j_{n}\left( \lambda \right) J_{n}\left(
\lambda r\right) }{\varkappa ^{2}+\beta _{\nu }^{2}}\left[ \cos \beta _{\nu
}z-\left( -1\right) ^{\nu }ch\left( \varkappa z\right) e^{-\varkappa l}%
\right] d\lambda -  \label{18}
\end{equation}%
\begin{equation*}
-\frac{\varepsilon _{n}}{2}\sum_{\nu =0}^{\infty }g_{nj}^{e+}J_{n}\left(
\gamma _{jn}\right) \int_{0}^{\infty }\frac{\lambda ^{2}J_{n}^{\shortmid
}\left( \lambda \right) J_{n}\left( \lambda r\right) }{\gamma
_{jn}^{2}-\lambda ^{2}}e^{-\varkappa l}sh\left( \shortmid \varkappa
z\shortmid \right) d\lambda -
\end{equation*}%
\begin{equation*}
-\frac{\varepsilon _{n}}{2}\sum_{\nu =0}^{\infty }g_{nj}^{\ast
e+}J_{n}\left( \gamma _{jn}\right) \int_{0}^{\infty }\frac{\lambda
^{2}J_{n}^{\shortmid }\left( \lambda \right) J_{n}\left( \lambda r\right) }{%
\varkappa \left( \gamma _{jn}^{2}-\lambda ^{2}\right) }e^{-\varkappa
l}sh\left( \shortmid \varkappa z\shortmid \right) d\lambda .
\end{equation*}

b)antisymmetric component
\begin{equation*}
P_{n}^{e-}\left( r,z\right) =\frac{\varepsilon _{n}}{2}\sum_{\nu =0}^{\infty
}f_{n\nu }^{e-}\int_{0}^{\infty }\frac{\lambda ^{2}j_{n}^{\shortmid }\left(
\lambda \right) J_{n}\left( \lambda r\right) }{\varkappa \left( \varkappa
^{2}+\beta _{\nu }^{2}\right) }\left[ \varkappa \sin \beta _{\nu }z-\left(
-1\right) ^{\nu }\beta _{\nu }sh\left( \varkappa z\right) e^{-\varkappa l}%
\right] d\lambda -
\end{equation*}%
\begin{equation*}
-\frac{\varepsilon _{n}}{2}\sum_{\nu =0}^{\infty }f_{n\nu }^{\ast
e-}\int_{0}^{\infty }\frac{\lambda J_{n}\left( \lambda \right) J_{n}\left(
\lambda r\right) }{\varkappa \left( \varkappa ^{2}+\beta _{\nu }^{2}\right) }%
\left[ \varkappa \sin \beta _{\nu }z-\left( -1\right) ^{\nu }\beta _{\nu
}sh\left( \varkappa z\right) e^{-\varkappa l}\right] d\lambda \pm
\end{equation*}%
\begin{equation}
\pm \frac{\varepsilon _{n}}{2}\sum_{\nu =0}^{\infty }g_{nj}^{e-}J_{n}\left(
\gamma _{jn}\right) \int_{0}^{\infty }\frac{\lambda ^{2}J_{n}^{\shortmid
}\left( \lambda \right) J_{n}\left( \lambda r\right) }{\gamma
_{jn}^{2}-\lambda ^{2}}e^{-\varkappa l}ch\left( \varkappa z\right) d\lambda
\pm  \label{19}
\end{equation}%
\begin{equation*}
\pm \frac{\varepsilon _{n}}{2}\sum_{\nu =0}^{\infty }g_{nj}^{\ast
e-}J_{n}\left( \gamma _{jn}\right) \int_{0}^{\infty }\frac{\lambda
^{2}J_{n}^{\shortmid }\left( \lambda \right) J_{n}\left( \lambda r\right) }{%
\varkappa \left( \gamma _{jn}^{2}-\lambda ^{2}\right) }e^{-\varkappa
l}sh\left( \shortmid \varkappa z\shortmid \right) d\lambda .
\end{equation*}%
where $\varepsilon _{0}=1;\varepsilon _{n}=2;n\geq 1.$

For the region $\left( r\geq 0,\left\vert z\right\vert \geq l,0\leq \Theta
\leq 2\pi \right) $ expression for $P_{n}^{e+}\left( r,z\right) $ and $%
P_{n}^{e-}\left( r,z\right) $ will look
\begin{equation}
P_{n}^{e+}\left( r,z\right) =\frac{\varepsilon _{n}}{2}\left\{
\begin{array}{c}
\sum_{\nu =0}^{\infty }f_{n\nu }^{e+}\int_{0}^{\infty }\frac{\lambda
^{2}J_{n}^{\shortmid }\left( \lambda \right) J_{n}\left( \lambda r\right) }{%
\varkappa ^{2}+\beta _{\nu }^{2}}\left( -1\right) ^{\nu }sh\left( \varkappa
l\right) e^{-\varkappa \mid z\mid }d\lambda - \\
-\sum_{\nu =0}^{\infty }f_{n\nu }^{\ast e+}\int_{0}^{\infty }\frac{\lambda
J_{n}\left( \lambda \right) J_{n}\left( \lambda r\right) }{\varkappa
^{2}+\beta _{\nu }^{2}}\left( -1\right) ^{\nu }sh\left( \varkappa l\right)
e^{-\varkappa \mid z\mid }d\lambda + \\
+\sum_{j=0}^{\infty }g_{nj}^{e+}J_{n}\left( \gamma _{jn}\right)
\int_{0}^{\infty }\frac{\lambda ^{2}J_{n}^{\shortmid }\left( \lambda \right)
J_{n}\left( \lambda r\right) }{\gamma _{jn}{}^{2}-\lambda ^{2}}sh\left(
\varkappa l\right) e^{-\varkappa \mid z\mid }d\lambda - \\
-\sum_{j=0}^{\infty }g_{nj}^{\ast e+}J_{n}\left( \gamma _{jn}\right)
\int_{0}^{\infty }\frac{\lambda ^{2}J_{n}^{\shortmid }\left( \lambda \right)
J_{n}\left( \lambda r\right) }{\varkappa \left( \gamma _{jn}{}^{2}-\lambda
^{2}\right) }ch\left( \varkappa l\right) e^{-\varkappa \mid z\mid }d\lambda%
\end{array}%
\right\}  \label{20}
\end{equation}%
\begin{equation}
P_{n}^{e-}\left( r,z\right) =\pm \frac{\varepsilon _{n}}{2}\left\{
\begin{array}{c}
-\sum_{\nu =0}^{\infty }f_{n\nu }^{e-}\int_{0}^{\infty }\frac{\lambda
^{2}J_{n}^{\shortmid }\left( \lambda \right) J_{n}\left( \lambda r\right) }{%
\varkappa ^{2}+\beta _{\nu }^{2}}\left( -1\right) ^{\nu }\beta _{\nu
}sh\left( \varkappa l\right) e^{-\varkappa \mid z\mid }d\lambda - \\
-\sum_{\nu =0}^{\infty }f_{n\nu }^{\ast e-}\int_{0}^{\infty }\frac{\lambda
J_{n}\left( \lambda \right) J_{n}\left( \lambda r\right) }{\varkappa \left(
\varkappa ^{2}+\beta _{\nu }^{2}\right) }\left( -1\right) ^{\nu }\beta _{\nu
}sh\left( \varkappa l\right) e^{-\varkappa \mid z\mid }d\lambda + \\
+\sum_{j=0}^{\infty }g_{nj}^{e-}J_{n}\left( \gamma _{jn}\right)
\int_{0}^{\infty }\frac{\lambda ^{2}J_{n}^{\shortmid }\left( \lambda \right)
J_{n}\left( \lambda r\right) }{\gamma _{jn}{}^{2}-\lambda ^{2}}ch\left(
\varkappa l\right) e^{-\varkappa \mid z\mid }d\lambda - \\
-\sum_{j=0}^{\infty }g_{nj}^{\ast e-}J_{n}\left( \gamma _{jn}\right)
\int_{0}^{\infty }\frac{\lambda ^{2}J_{n}^{\shortmid }\left( \lambda \right)
J_{n}\left( \lambda r\right) }{\varkappa \left( \gamma _{jn}{}^{2}-\lambda
^{2}\right) }sh\left( \varkappa l\right) e^{-\varkappa \mid z\mid }d\lambda%
\end{array}%
\right\}  \label{21}
\end{equation}

+ when $z>l$

- when $z<-l.$

Functions $P_{n}^{e\pm }\left( r,z\right) $ and $P_{n}^{e\pm }\left(
r_{0},z_{0}\right) $ and introduced in series (\ref{16})-(\ref{17}) unknown
coefficients $f_{n\nu }^{e\pm },g_{nj}^{e\pm }$ and functions $P_{n}^{e\pm
}\left( r_{0},z_{0}\right) $ are tied by relations \cite{10, 11, 12}
\begin{equation*}
f_{n\nu }^{e+}=\frac{\varepsilon _{\nu }}{l}\int_{-j}^{l}P_{n}^{e+}\left(
r,z\right) \mid _{r=1}\cos \left( \beta _{\nu }z\right) dz;\varepsilon _{0}=%
\frac{1}{2};\varepsilon _{\nu }=1,\nu \geq 1;
\end{equation*}%
\begin{equation}
f_{n\nu }^{e-}=\frac{1}{l}\int_{-j}^{l}P_{n}^{e-}\left( r,z\right) \mid
_{r=1}\sin \left( \beta _{\nu }z\right) dz;  \label{22}
\end{equation}%
\begin{equation*}
g_{nj}^{e\pm }=\varepsilon _{nj}\int_{0}^{1}P_{n}^{e\pm }\left( r,z\right)
\mid _{z=l}J_{n}\left( \gamma _{jn}r\right) rdr;\epsilon _{nj}=\frac{2\gamma
_{jn}^{2}}{\left( \gamma _{jn}^{2}-n^{2}\right) J_{n}^{2}\left( \gamma
_{jn}\right) }.
\end{equation*}

Substituting the expression (\ref{18}),(\ref{19}) into (\ref{22}) we get
algebraic equations
\begin{equation*}
\sum_{\nu =0}^{\infty }\left( f_{n\nu }^{e\pm }F_{n\nu \mu }^{1\pm }+f_{n\nu
}^{\ast e\pm }F_{n\nu \mu }^{\ast 1\pm }\right) +\sum_{j=0}^{\infty }\left(
g_{nj}^{e\pm }G_{nj\mu }^{1\pm }+g_{nj}^{\ast e\pm }G_{nj\mu }^{\ast 1\pm
}\right) =0;
\end{equation*}%
\begin{equation}
\sum_{\nu =0}^{\infty }\left( f_{n\nu }^{e\pm }F_{n\nu \mu }^{2\pm }+f_{n\nu
}^{\ast e\pm }F_{n\nu \mu }^{\ast 2\pm }\right) +\sum_{j=0}^{\infty }\left(
g_{nj}^{e\pm }G_{nj\mu }^{2\pm }+g_{nj}^{\ast e\pm }G_{nj\mu }^{\ast 2\pm
}\right) =0.  \label{23}
\end{equation}

where
\begin{equation*}
F_{N\nu \mu }^{\ast 1+}=-\int_{0}^{\infty }\frac{\lambda J_{n}^{2}\left(
\lambda \right) }{\left( \varkappa ^{2}+\beta _{\nu }^{2}\right) }\left[
\varepsilon _{\mu }l\delta _{\nu \mu }-\frac{\left( -1\right) ^{\nu +\mu
}\varkappa \left( 1-e^{-2\varkappa l}\right) }{\left( \varkappa ^{2}+\beta
_{\nu }^{2}\right) }\right] d\lambda ;
\end{equation*}%
\begin{equation*}
F_{N\nu \mu }^{1+}=\int_{0}^{\infty }\frac{J_{n}^{\shortmid }\left( \lambda
\right) J_{n}\left( \lambda \right) }{\left( \varkappa ^{2}+\beta _{\nu
}^{2}\right) }\left[ \varepsilon _{\mu }l\delta _{\nu \mu }\left( \omega
^{2}-\beta _{\nu }^{2}\right) -\frac{\left( -1\right) ^{\nu +\mu }\left(
y_{1}-y_{2}\right) }{\varkappa \left( \varkappa ^{2}+\beta _{\nu
}^{2}\right) }\right] d\lambda -
\end{equation*}%
\begin{equation*}
-\varepsilon _{\mu }l\delta _{\nu \mu }\left( \frac{n}{2}+1\right)
;y_{1}=\varkappa ^{2}\lambda ^{2}\left( 1-e^{-2\varkappa l}\right)
;y_{2}=\left( \varkappa ^{2}+\beta _{\nu }^{2}\right) ^{2};
\end{equation*}%
\begin{equation*}
G_{nj\mu }^{1+}=J_{n}\left( \gamma _{jn}\right) \int_{0}^{\infty }\frac{%
\lambda ^{2}J_{n}^{\shortmid }\left( \lambda \right) J_{n}\left( \lambda
\right) \left( -1\right) ^{\mu }\left( 1-e^{-2\varkappa l}\right) }{\left(
\gamma _{jn}^{2}-\lambda ^{2}\right) \left( \varkappa ^{2}+\beta _{\mu
}^{2}\right) }d\lambda
\end{equation*}%
$\ \ \ \ \ \ \ \ \ \ \ \ \ \ \ \ \ \ \ \ \ \ \ \ \ \ \ \ \ \ \ \ $%
\begin{equation*}
G_{nj\mu }^{\ast 1+}=-J_{n}\left( \gamma _{jn}\right) \int_{0}^{\infty }%
\frac{\lambda ^{2}J_{n}^{\shortmid }\left( \lambda \right) J_{n}\left(
\lambda \right) \left( -1\right) ^{\mu }\varkappa \left( 1-e^{-2\varkappa
l}\right) }{\left( \gamma _{jn}^{2}-\lambda ^{2}\right) \left( \varkappa
^{2}+\beta _{\mu }^{2}\right) }d\lambda ;
\end{equation*}%
\begin{equation*}
G_{nj\mu }^{2+}=J_{n}\left( \gamma _{jn}\right) J_{n}\left( \gamma _{\mu
n}\right) \int_{0}^{\infty }\frac{\lambda ^{3}J_{n}^{\shortmid 2}\left(
\lambda \right) \left( 1-e^{-2\varkappa l}\right) }{\left( \gamma
_{jn}^{2}-\lambda ^{2}\right) \left( \gamma _{\mu n}^{2}-\lambda ^{2}\right)
}d\lambda -\frac{\delta _{j\mu }}{\varepsilon _{n\mu }};
\end{equation*}%
\begin{equation*}
G_{nj\mu }^{\ast 2+}=J_{n}\left( \gamma _{jn}\right) J_{n}\left( \gamma
_{\mu n}\right) \int_{0}^{\infty }\frac{\lambda ^{3}J_{n}^{\shortmid
2}\left( \lambda \right) \left( 1+e^{-2\varkappa l}\right) }{\varkappa
\left( \gamma _{jn}^{2}-\lambda ^{2}\right) \left( \gamma _{\mu
n}^{2}-\lambda ^{2}\right) }d\lambda ;
\end{equation*}%
\begin{equation*}
F_{n\nu \mu }^{2+}=J_{n}\left( \gamma _{\mu n}\right) \int_{0}^{\infty }%
\frac{\lambda ^{3}J_{n}^{\shortmid 2}\left( \lambda \right) \left( -1\right)
^{\nu }\left( 1-e^{-2\varkappa l}\right) }{\left( \varkappa ^{2}+\beta _{\mu
}^{2}\right) \left( \gamma _{\mu n}^{2}-\lambda ^{2}\right) }d\lambda ;
\end{equation*}%
\begin{equation*}
F_{n\nu \mu }^{\ast 2+}=J_{n}\left( \gamma _{\mu n}\right) \int_{0}^{\infty }%
\frac{\lambda ^{2}J_{n}^{\shortmid }\left( \lambda \right) J_{n}\left(
\lambda \right) \left( -1\right) ^{\nu }\left( 1-e^{-2\varkappa l}\right) }{%
\left( \varkappa ^{2}+\beta _{\mu }^{2}\right) \left( \gamma _{\mu
n}^{2}-\lambda ^{2}\right) }d\lambda ;
\end{equation*}%
\begin{equation*}
F_{n\nu \mu }^{1-}=\int_{0}^{\infty }\frac{J_{n}^{\shortmid }\left( \lambda
\right) J_{n}\left( \lambda \right) }{\varkappa \left( \varkappa ^{2}+\beta
_{\nu }^{2}\right) }\left[ \varkappa l\delta _{\nu \mu }\left( \varpi
^{2}-\beta _{\nu }^{2}\right) -\frac{\left( -1\right) ^{\nu +\mu }\beta
_{\nu }\beta _{\mu }\lambda ^{2}\left( 1-e^{-2\varkappa l}\right) }{\left(
\varkappa ^{2}+\beta _{\mu }^{2}\right) }\right] d\lambda -
\end{equation*}%
\begin{equation*}
-\delta _{\nu \mu }l\left( \frac{n}{2}+1\right) ;
\end{equation*}%
\begin{equation*}
F_{n\nu \mu }^{\ast 1-}=\int_{0}^{\infty }\frac{\lambda J_{n}^{2}\left(
\lambda \right) }{\varkappa \left( \varkappa ^{2}+\beta _{\nu }^{2}\right) }%
\left[ \varkappa l\delta _{\nu \mu }+\frac{\left( -1\right) ^{\nu +\mu
}\beta _{\nu }\left( 1-e^{-2\varkappa l}\right) }{\left( \varkappa
^{2}+\beta _{\mu }^{2}\right) }\right] d\lambda ;
\end{equation*}%
\begin{equation*}
G_{nj\mu }^{1-}=J_{n}\left( \gamma _{jn}\right) \int_{0}^{\infty }\frac{%
\lambda ^{2}J_{n}^{\shortmid }\left( \lambda \right) J_{n}\left( \lambda
\right) \left( -1\right) ^{\mu }\beta _{\mu }\left( 1-e^{-2\varkappa
l}\right) }{\left( \gamma _{jn}^{2}-\lambda ^{2}\right) \left( \varkappa
^{2}+\beta _{\mu }^{2}\right) }d\lambda ;
\end{equation*}%
\begin{equation*}
G_{nj\mu }^{\ast 1-}=J_{n}\left( \gamma _{jn}\right) \int_{0}^{\infty }\frac{%
\lambda ^{2}J_{n}^{\shortmid }\left( \lambda \right) J_{n}\left( \lambda
\right) \left( -1\right) ^{\mu }\beta _{\mu }\left( 1-e^{-2\varkappa
l}\right) }{\varkappa \left( \gamma _{jn}^{2}-\lambda ^{2}\right) \left(
\varkappa ^{2}+\beta _{\mu }^{2}\right) }d\lambda ;
\end{equation*}%
\begin{equation*}
F_{n\nu \mu }^{2-}=-J_{n}\left( \gamma _{\mu n}\right) \int_{0}^{\infty }%
\frac{\lambda ^{3}J_{n}^{\shortmid }\left( \lambda \right) \left( -1\right)
^{\nu }\beta _{\nu }\left( 1-e^{-2\varkappa l}\right) }{\varkappa \left(
\varkappa ^{2}+\beta _{\mu }^{2}\right) \left( \gamma _{\mu n}^{2}-\lambda
^{2}\right) }d\lambda ;
\end{equation*}%
\begin{equation*}
F_{nj\mu }^{\ast 2-}=-J_{n}\left( \gamma _{\mu n}\right) \int_{0}^{\infty }%
\frac{\lambda ^{2}J_{n}^{\shortmid }\left( \lambda \right) J_{n}\left(
\lambda \right) \left( -1\right) ^{\mu }\beta _{\nu }\left( 1-e^{-2\varkappa
l}\right) }{\varkappa \left( \gamma _{jn}^{2}-\lambda ^{2}\right) \left(
\varkappa ^{2}+\beta _{\mu }^{2}\right) }d\lambda ;
\end{equation*}%
\begin{equation*}
G_{nj\mu }^{2-}=J_{n}\left( \gamma _{jn}\right) J_{n}\left( \gamma _{\mu
n}\right) \int_{0}^{\infty }\frac{\lambda ^{3}J_{n}^{\shortmid 2}\left(
\lambda \right) \left( 1+e^{-2\varkappa l}\right) }{\left( \gamma
_{jn}^{2}-\lambda ^{2}\right) \left( \gamma _{\mu n}^{2}-\lambda ^{2}\right)
}d\lambda -\frac{\delta _{j\mu }}{\varepsilon _{n\mu }};
\end{equation*}%
\begin{equation*}
G_{nj\mu }^{\ast 2-}=J_{n}\left( \gamma _{jn}\right) J_{n}\left( \gamma
_{\mu n}\right) \int_{0}^{\infty }\frac{\lambda ^{3}J_{n}^{\shortmid
2}\left( \lambda \right) \left( 1-e^{-2\varkappa l}\right) }{\varkappa
\left( \gamma _{jn}^{2}-\lambda ^{2}\right) \left( \gamma _{\mu
n}^{2}-\lambda ^{2}\right) }d\lambda .
\end{equation*}%
$\delta _{\nu \mu }$ is the Kronecker's symbol.

Let us consider a question of derivatives representation from series (\ref%
{13}) also by Fourier and Fourier-Bessel series, as during the solution of
equations (\ref{3})-(\ref{8}) we need to differentiate it.

Let us assume that even function $f\left( z\right) $ and odd function $%
\varphi \left( z\right) $ and its derivatives could be expanded into Fourier
series in $\left[ -l,l\right] :$%
\begin{equation*}
f^{\left( 2k\right) }\left( z\right) =\sum_{\nu =0}^{\infty }a_{\nu
}^{2k}\cos \beta _{\nu }z;\varphi ^{\left( 2k\right) }\left( z\right)
=\sum_{\nu =1}^{\infty }b_{\nu }^{2k}\sin \beta _{\nu }z;
\end{equation*}%
\begin{equation}
f^{\left( 2k+1\right) }\left( z\right) =\sum_{\nu =0}^{\infty }a_{\nu
}^{2k+1}\sin \beta _{\nu }z;\varphi ^{\left( 2k-1\right) }\left( z\right)
=\sum_{\nu =1}^{\infty }b_{\nu }^{2k-1}\cos \beta _{\nu }z.  \label{24}
\end{equation}%
Coefficients $a_{\nu }^{2K},a_{\nu }^{2K-1},b_{\nu }^{2K},b_{\nu }^{2K-1}$
are defined by formulas \cite{13, 14, 15}%
\begin{equation*}
a_{\nu }^{2k}=\frac{\varepsilon _{\nu }}{2}\int_{-l}^{l}f^{\left( 2k\right)
}\left( z\right) \cos \beta _{\nu }zdz;b_{\nu }^{2k}=\frac{\varepsilon _{\nu
}}{2}\int_{-l}^{l}\varphi ^{\left( 2k\right) }\left( z\right) \sin \beta
_{\nu }zdz
\end{equation*}%
\begin{equation}
a_{\nu }^{2k+1}=\frac{\varepsilon _{\nu }}{2}\int_{-l}^{l}f^{\left(
2k+1\right) }\left( z\right) \sin \beta _{\nu }zdz;b_{\nu }^{2k-1}=\frac{%
\varepsilon _{\nu }}{2}\int_{-l}^{l}\varphi ^{\left( 2k-1\right) }\left(
z\right) \cos \beta _{\nu }zdz  \label{25}
\end{equation}%
Integrating (\ref{25}) by parts and taking into account the evenness and
oddness of derivatives from $f\left( z\right) $ and $\varphi \left( z\right)
$ we find that
\begin{equation*}
a_{\nu }^{2k}=\frac{2\varepsilon _{\nu }}{l}\sum \left[ \left( -1\right)
^{i+\nu +1}\beta _{\nu }^{2i-2}f^{\left( 2k-2i+1\right) }\left( l\right)
+\left( -1\right) ^{k}\beta _{\nu }^{2k}a_{\nu }^{0}\right] ;
\end{equation*}%
\begin{equation}
b_{\nu }^{2k-1}=\frac{2}{l}\sum \left[ \left( -1\right) ^{i+\nu +1}\beta
_{\nu }^{2i-2}\varphi ^{\left( 2k-2i\right) }\left( l\right) -\left(
-1\right) ^{k}\beta _{\nu }^{2k-1}b_{\nu }^{0}\right] ;  \label{26}
\end{equation}%
\begin{equation*}
a_{\nu }^{2k+1}=\beta _{\nu }a_{\nu }^{2k};b_{\nu }^{2k}=-\beta _{\nu
}b_{\nu }^{2k-1};
\end{equation*}

During the solution of equations (\ref{3}) we present the function $%
P_{n}^{0}\left( r,z\right) $ as the series
\begin{equation}
\sum_{\nu =0}^{\infty }\left[ P_{n\nu }^{o+}\left( r\right) \cos \beta _{\nu
}z+P_{n\nu }^{o-}\left( r\right) \sin \beta _{\nu }z\right] =P_{n}^{o}\left(
r,z\right)  \label{27}
\end{equation}

The second derivative from $P_{n}^{0}\left( r,z\right) $ taking into account
(\ref{24}), (\ref{26}) we write as
\begin{equation*}
\frac{\partial ^{2}}{\partial z^{2}}P_{n}^{o}\left( r,z\right) =\sum_{\nu
=0}^{\infty }\left( \frac{\partial }{\partial z}P_{n}^{o+}\left( r,z\right)
\mid _{z=l}\frac{2\varepsilon _{\nu }\left( -1\right) ^{\nu }}{l}-\beta
_{\nu }^{2}P_{n\nu }^{o+}\left( r\right) \right) \cos \beta _{\nu }z+
\end{equation*}%
\begin{equation}
+\sum_{\nu =0}^{\infty }\left( -\frac{2\left( -1\right) ^{\nu }\beta _{\nu }%
}{l}P_{n}^{o+}\left( r,z\right) \mid _{z=l}-\beta _{\nu }^{2}P_{n\nu
}^{o-}\left( r\right) \right) \sin \beta _{\nu }z  \label{28}
\end{equation}

Let the dissipated pressure and its derivatives on the interior side of the
plates are defined as follows
\begin{equation}
P_{n}^{o\pm }\left( r,l\right) =\sum_{j=0}^{\infty }g_{nj}^{o\pm
}J_{n}\left( \gamma _{jn}r\right) ;\frac{\partial }{\partial z}P_{n}^{o\pm
}\left( r,l\right) =\sum_{j=0}^{\infty }g_{nj}^{\ast o\pm }J_{n}\left(
\gamma _{jn}r\right)  \label{29}
\end{equation}%
Taking into account (\ref{17})-(\ref{29}) wave equation (\ref{3}) for the
symmetrical and antisymmetric components obtains the following form
\begin{equation}
\left[ \frac{\partial ^{2}}{\partial r^{2}}+\frac{1}{r}\frac{\partial }{%
\partial r}-\left( \frac{n^{2}}{r^{2}}-\varkappa _{2}^{2}\right) \right]
P_{n\nu }^{o+}\left( r\right) =-\sum_{j=0}^{\infty }\frac{2\varepsilon _{\nu
}\left( -1\right) ^{\nu }}{l}g_{nj}^{\ast o+}J_{n}\left( \gamma
_{jn}r\right) ;  \label{30}
\end{equation}%
\begin{equation*}
\left[ \frac{\partial ^{2}}{\partial r^{2}}+\frac{1}{r}\frac{\partial }{%
\partial r}-\left( \frac{n^{2}}{r^{2}}-\varkappa _{2}^{2}\right) \right]
P_{n\nu }^{o-}\left( r\right) =\sum_{j=0}^{\infty }\frac{2\beta _{\nu
}\left( -1\right) ^{\nu }}{l}g_{nj}^{o-}J_{n}\left( \gamma _{jn}r\right) ;
\end{equation*}%
where \ae $_{2}^{2}=\frac{\omega ^{2}}{\beta _{0}^{2}-\beta _{\nu }^{2}};$

General solutions of (\ref{30}) are
\begin{equation}
P_{n\nu }^{o+}\left( r\right) =f_{n\nu }^{o+}J_{n}\left( \varkappa
_{2}r\right) -\sum_{j=0}^{\infty }C_{n\nu j}^{+}g_{nj}^{\ast o+}J_{n}\left(
\gamma _{jn}r\right) ;  \label{31}
\end{equation}%
\begin{equation*}
P_{n\nu }^{o-}\left( r\right) =f_{n\nu }^{o-}J_{n}\left( \varkappa
_{2}r\right) +\sum_{j=0}^{\infty }C_{n\nu j}^{-}g_{nj}^{o-}J_{n}\left(
\gamma _{jn}r\right) ;
\end{equation*}%
where
\begin{equation*}
C_{n\nu j}^{+}=\frac{2\varepsilon _{\nu }\left( -1\right) ^{\nu }}{l\left(
\varkappa _{2}^{2}-\gamma _{jn}^{2}\right) };C_{n\nu j}^{-}=\frac{2\beta
_{\nu }\left( -1\right) ^{\nu }}{l\left( \varkappa _{2}^{2}-\gamma
_{jn}^{2}\right) }
\end{equation*}

From (\ref{31}) we obtain that pressure $P_{n}^{0\pm }\left( r\right) $ and $%
\frac{\partial }{\partial r}P_{n}^{0\pm }\left( r\right) $ on the interior
surface of cylindrical cover $\left( z=1\right) $ are
\begin{eqnarray}
\frac{\partial }{\partial r}P_{n}^{o\pm }\left( 1\right) &=&f_{n\nu }^{o\pm
}\left( \varkappa _{2}\right) J_{n}^{\shortmid }\left( \varkappa _{2}\right)
.  \label{32} \\
P_{n\nu }^{o+}\left( 1\right) &=&f_{n\nu }^{o+}J_{n}\left( \varkappa
_{2}\right) -\sum_{j=0}^{\infty }C_{n\nu j}^{+}g_{nj}^{\ast o+}J_{n}\left(
\gamma _{jn}\right)  \notag
\end{eqnarray}%
\begin{equation*}
P_{n\nu }^{o-}\left( 1\right) =f_{n\nu }^{o-}J_{n}\left( \varkappa
_{2}\right) +\sum_{j=0}^{\infty }C_{n\nu j}^{-}g_{nj}^{o-}J_{n}\left( \gamma
_{jn}\right)
\end{equation*}

Let us exemplify $P_{n\nu }^{0\pm }$ from (\ref{31}) by Fourier-Bessel's
series and equate it to coefficients (\ref{29}). We get
\begin{equation}
g_{nj}^{o-}=\sum_{\nu =0}^{\infty }f_{n\nu }^{o+}\frac{\left( -1\right)
^{\nu }\varkappa _{2}J_{n}^{\shortmid }\left( \varkappa _{2}\right)
J_{n}\left( \gamma _{jn}\right) \varepsilon _{n\mu }}{\gamma
_{jn}^{2}-\varkappa _{2}^{2}}-g_{nj}^{\ast o+}\sum_{\nu =0}^{\infty }\left(
-1\right) ^{\nu }C_{n\nu \mu }^{+};  \label{33}
\end{equation}%
\begin{equation}
g_{nj}^{\ast o-}=\sum_{\nu =0}^{\infty }f_{n\nu }^{o+}\frac{\left( -1\right)
^{\nu }\varkappa _{2}J_{n}^{\shortmid }\left( \varkappa _{2}\right)
J_{n}\left( \gamma _{jn}\right) \varepsilon _{n\mu }}{\gamma
_{jn}^{2}-\varkappa _{2}^{2}}-g_{nj}^{o+}\sum_{\nu =0}^{\infty }\left(
-1\right) ^{\nu }C_{n\nu \mu }^{-}.  \label{34}
\end{equation}

Let us represent the falling pressure $P^{a}\left( r,z,\Theta \right) $ and
ints normal derivatives on the surface of a cover and plates a the series
analogical to (\ref{13}), (\ref{17}), (\ref{18}):
\begin{equation*}
P^{a}\left( r,z,\theta \right) =\sum_{n=0}^{\infty }P_{n}^{a}\left(
r,z\right) \cos n\theta ;
\end{equation*}%
\begin{equation*}
P_{n}^{a}\left( 1,z\right) =\sum_{\nu =0}^{\infty }\left( f_{n\nu }^{a+}\cos
\beta _{\nu }z+f_{n\nu }^{a-}\sin \beta _{\nu }z\right) ;P_{n}^{a\pm
z}\left( r,l\right) =\sum_{j=0}^{\infty }g_{nj}^{a\pm }J_{n}\left( \gamma
_{jn}r\right) ;
\end{equation*}%
\begin{equation}
\frac{\partial }{\partial r}P_{n}^{a}\left( r,z\right) \mid _{r=1}=\sum_{\nu
=0}^{\infty }\left( f_{n\nu }^{\ast a+}\cos \beta _{\nu }z+f_{n\nu }^{\ast
a-}\sin \beta _{\nu }z\right) ;  \label{35}
\end{equation}%
\begin{equation*}
\frac{\partial }{\partial z}P_{n}^{a\pm }\left( r,l\right) \mid _{z=\pm
j}=\sum_{j=0}^{\infty }\left( g_{nj}^{\ast a+}\pm g_{nj}^{\ast a+}\right)
J_{n}\left( \gamma _{jn}r\right) ;
\end{equation*}

Let us present the flat wave (\ref{1}) as the expansion \cite{8, 21}%
\begin{equation*}
P^{a}\left( r,z,\theta \right) =P_{0}\sum_{j=0}^{\infty }\varepsilon
_{n}i^{-n}J_{n}\left( \omega \sin \varphi ^{\ast }r\right) \exp \left(
-i\omega z\cos \varphi ^{\ast }\right) \cos n\theta ;
\end{equation*}%
and integrating using formulas that define the correspondent series
components (\ref{35}) we obtain
\begin{equation*}
f_{n\nu }^{a+}=-\frac{\left( -1\right) ^{\nu }2\varepsilon
_{n}i^{-n}J_{n}\left( \omega \sin \varphi ^{\ast }\right) \omega \cos
\varphi ^{\ast }}{\beta _{\nu }^{2}-\omega ^{2}\cos ^{2}\varphi ^{\ast }}%
\sin \left( \omega l\cos \varphi ^{\ast }\right) ;
\end{equation*}%
\begin{equation*}
f_{n\nu }^{a-}=\frac{2\left( -1\right) ^{\nu }\varepsilon
_{n}i^{-n}J_{n}\left( \omega \sin \varphi ^{\ast }\right) \beta _{\nu }}{%
\beta _{\nu }^{2}-\omega ^{2}\cos ^{2}\varphi ^{\ast }}\sin \left( \omega
l\cos \varphi ^{\ast }\right) ;
\end{equation*}%
\begin{equation*}
g_{n\nu }^{a-}=-\frac{\varepsilon _{nj}\varepsilon _{n}i^{-n}\omega \sin
\varphi ^{\ast }J_{n}^{\shortmid }\left( \omega \sin \varphi ^{\ast }\right)
J_{n}\left( \gamma _{jn}\right) }{\gamma _{jn}^{2}-\omega ^{2}\sin
^{2}\varphi ^{\ast }}\cos \left( \omega l\cos \varphi ^{\ast }\right) ;
\end{equation*}%
\begin{equation*}
g_{n\nu }^{a+}=\frac{\varepsilon _{n}i^{-n}\varepsilon _{nj}\left( \omega
\sin \varphi ^{\ast }\right) J_{n}\left( \gamma _{jn}\right) }{\gamma
_{jn}^{2}-\omega ^{2}\sin ^{2}\varphi ^{\ast }}\sin \left( \omega l\cos
\varphi ^{\ast }\right) ;
\end{equation*}%
\begin{equation*}
f_{n\nu }^{\ast a+}=-\varepsilon _{n}i^{-n}\omega \sin \varphi ^{\ast
}J_{n}^{\shortmid }\left( \omega \sin \varphi ^{\ast }\right) \frac{2\cos
\varphi ^{\ast }\left( -1\right) ^{\nu }\omega }{\beta _{\nu }^{2}-\omega
^{2}\cos ^{2}\varphi ^{\ast }}\sin \left( \omega l\cos \varphi ^{\ast
}\right) ;
\end{equation*}%
\begin{equation*}
f_{n\nu }^{\ast a-}=\varepsilon _{n}i^{-n}\omega \sin \varphi ^{\ast
}J_{n}^{\shortmid }\left( \omega \sin \varphi ^{\ast }\right) \frac{2\beta
_{\nu }}{\beta _{\nu }^{2}-\omega ^{2}\cos ^{2}\varphi ^{\ast }}\sin \left(
\omega l\cos \varphi ^{\ast }\right) ;
\end{equation*}%
\begin{equation*}
g_{n\nu }^{\ast a+}=i\omega \cos \varphi ^{\ast }g_{nj}^{a+};g_{n\nu }^{\ast
a-}=i\omega \cos \varphi ^{\ast }g_{nj}^{a-}.
\end{equation*}

To find the solution of inhomogeneous equation (\ref{5}) we define the
auxiliary function
\begin{equation}
F^{\pm }=\Delta _{0}W^{\pm }+\frac{S}{D_{2}}g_{z}^{\pm };  \label{36}
\end{equation}

Taking into account (\ref{36}) to define $F^{\pm }\left( r,\Theta \right) $
from (\ref{5}) we get the following differential equation:
\begin{equation}
\left( \Delta _{0}^{2}-\alpha _{1}\Delta _{0}-\alpha _{2}\right) F^{\pm
}\left( r,\theta \right) =\left( \beta _{1}^{\ast }-\beta _{2}^{\ast
}\right) g_{z}^{\pm }\left( r,z\right) ;  \label{37}
\end{equation}%
where
\begin{equation*}
\beta _{1}^{\ast }=\alpha _{2}^{\ast }+\alpha _{1}^{\ast }\frac{S}{D_{2}}%
;\beta _{2}^{\ast }=\alpha _{2}\frac{S}{D_{2}};
\end{equation*}%
\begin{equation*}
g_{z}^{\pm }\left( r,z\right) =\sum_{n=0}^{\infty }\sum_{j=0}^{\infty
}\left( g_{nj}^{a\pm }+g_{nj}^{e\pm }-g_{nj}^{o\pm }\right) J_{n}\left(
\gamma _{jn}r\right) \cos n\theta ;
\end{equation*}%
The solution of equation (\ref{37}) we write as follows:
\begin{equation*}
F^{\pm }\left( r,\theta \right) =\sum_{n=0}^{\infty }F_{n}^{\pm }\left(
r\right) \cos n\theta
\end{equation*}%
\begin{equation}
F_{n}^{\pm }\left( r\right) =A_{1n}^{\pm }J_{n}\left( \gamma _{1}r\right)
+A_{2n}^{\pm }J_{n}\left( \gamma _{2}r\right) +\sum_{j=0}^{\infty
}A_{3nj}g_{znj}J_{n}\left( \gamma _{jn}r\right)  \label{38}
\end{equation}%
\begin{equation*}
\gamma _{1,2}=\sqrt{\frac{\alpha _{1}\pm \sqrt{\alpha _{1}^{2}+4\alpha _{2}}%
}{2}};A_{3nj}=\frac{-\beta _{1}^{\ast }\gamma _{jn}^{2}-\beta _{2}^{\ast }}{%
\gamma _{jn}^{4}-\alpha _{1}\gamma _{jn}^{2}-\alpha _{2}};
\end{equation*}%
\begin{equation*}
g_{znj}^{\pm }=g_{nj}^{a\pm }+g_{nj}^{e\pm }+g_{nj}^{o.\pm }.
\end{equation*}%
$A_{1n}^{\pm },A_{2n}^{\pm }$ are unknown coefficients.

Using (\ref{38}) and (\ref{36}) to the normal shifts of the plates we obtain
\begin{equation}
W_{n}^{\pm }\left( r\right) =\frac{A_{1n}^{\pm }J_{n}\left( \gamma
_{1}r\right) }{-\gamma _{1}^{2}}+\frac{A_{2n}^{\pm }J_{n}\left( \gamma
_{2}r\right) }{-\gamma _{2}^{2}}+\sum_{j=0}^{\infty }\frac{\left( A_{3nj}-%
\frac{S}{D_{2}}\right) }{-\gamma _{jn}^{2}}g_{znj}^{\pm }J_{n}\left( \gamma
_{jn}r\right) .  \label{39}
\end{equation}%
The solution of equations (\ref{6})-(\ref{8}) will be rewritten as follows:
\begin{equation}
\Phi _{n}^{\pm }\left( r\right) =\Phi _{1n}^{\pm }J_{n}\left( k_{1}r\right) ;
\label{40}
\end{equation}%
\begin{equation}
\varphi _{n}^{\pm }\left( r\right) =\varphi _{1n}^{\pm }J_{n}\left(
k_{1}r\right)  \label{41}
\end{equation}%
\begin{equation}
\Psi _{n}^{\pm }\left( r\right) =\Psi _{1n}^{\pm }J_{n}\left( k_{2}r\right)
\label{42}
\end{equation}

Substituting expansions (\ref{13}) into equation (\ref{4}) taking into
account (\ref{24}), (\ref{26}) we get the algebraic equations relatively to
components of shift vector components:
\begin{equation}
a_{ij}^{\pm }u_{nj\nu }=b_{in}^{\pm }+\delta _{i3}g_{n\nu }^{\pm
};a_{ij}^{\pm }=a_{ji}^{\pm };i,j=1,2,3,4,5.  \label{43}
\end{equation}%
where
\begin{equation*}
a_{11}^{\pm }=\left( 1+a^{2}\right) \left( -n^{2}-\sigma _{1}\beta _{\nu
}^{2}+\varkappa _{1}^{2}\right) -\sigma _{1}\beta _{\nu }^{2}-\frac{\omega
^{2}}{\beta _{\nu }^{2}};
\end{equation*}%
\begin{equation*}
a_{12}^{\pm }=a^{2}\left( n^{2}-\sigma _{1}\beta _{\nu }^{2}+\varkappa
_{1}^{2}\right) -\frac{\omega ^{2}}{\beta _{\nu }^{2}};a_{13}^{\pm }=\left(
1+a^{2}\right) \left( 1+\varkappa _{1}^{2}\right) \left( \pm n\right) ;
\end{equation*}%
\begin{equation*}
a_{14}^{\pm }=\beta _{\nu }\sigma _{2}\left( _{+}^{-}n\right) ;a_{15}^{\pm
}=0;a_{22}^{\pm }=a^{2}\left( -n^{2}-\sigma _{1}\beta _{\nu }^{2}-\varkappa
_{1}^{2}+\frac{\omega ^{2}}{\beta _{\nu }^{2}}\right) -\varkappa _{1}^{2};
\end{equation*}%
\begin{equation*}
a_{23}^{\pm }=\left[ \left( 1+a^{2}\right) \varkappa _{1}^{2}+a^{2}\right]
\left( \pm n\right) ;a_{24}^{\pm }=0;a_{25}^{\pm }=a^{2}\sigma _{2}\beta
_{\nu }^{2}\left( \pm n\right) ;
\end{equation*}%
\begin{equation*}
a_{34}^{\pm }=\nu _{01}\beta _{\nu };a_{33}^{\pm }=\left( 1+a^{2}\right)
\left( 1+\varkappa _{1}^{2}n^{2}\right) +\varkappa _{1}^{2}\beta _{\nu }^{2}-%
\frac{\omega ^{2}}{\beta _{\nu }^{2}};a_{35}^{\pm }=_{+}^{-}\beta _{\nu
}\varkappa _{1}^{2}
\end{equation*}%
\begin{equation*}
a_{44}^{\pm }=-\left( 1+a^{2}\right) \sigma _{1}n^{2}-\beta _{\nu }^{2}+%
\frac{\omega ^{2}}{\beta _{\nu }^{2}};a_{45}^{\pm }=a^{2}\left( n^{2}\sigma
_{1}-\beta _{\nu }^{2}+\frac{\omega ^{2}}{\beta _{\nu }^{2}}\right) ;
\end{equation*}%
\begin{equation*}
a_{55}^{\pm }=a^{2}\left( -n^{2}\sigma _{1}-\beta _{\nu }^{2}+\frac{\omega
^{2}}{\beta _{\nu }^{2}}\right) ;
\end{equation*}%
\begin{equation*}
b_{1n}^{+}=C_{11}^{+}u_{1n}^{\shortmid +}+C_{12}^{+}u_{2n}^{\shortmid
+}+C_{14}^{+}u_{4n}^{+};b_{2n}^{+}=C_{21}^{+}u_{1n}^{\shortmid
+}+C_{22}^{+}u_{2n}^{\shortmid +}+C_{25}^{+}u_{5n}^{+};
\end{equation*}%
\begin{equation*}
b_{3n}^{+}=C_{33}^{+}u_{3n}^{\shortmid
+}+C_{34}^{+}u_{4n}^{+}+C_{35}^{+}u_{5n}^{+}+g_{rn\nu
}^{+};b_{4n}^{+}=C_{44}^{+}u_{4n}^{+}+C_{45}^{+}u_{5n}^{+};
\end{equation*}%
\begin{equation*}
b_{5n}^{+}=C_{54}^{+}u_{4n}^{+}+C_{55}^{+}u_{5n}^{+};b_{1n}^{-}=C_{11}^{-}u_{1n}^{-}+C_{12}^{-}u_{2n}^{-}+C_{14}^{-}u_{4n}^{-};
\end{equation*}%
\begin{equation*}
b_{2n}^{-}=C_{21}^{-}u_{1n}^{-}+C_{22}^{-}u_{2n}^{-}+C_{25}^{-}u_{5n}^{-};b_{3n}^{-}=C_{33}^{-}u_{3n}^{-}+g_{rn\nu }^{-};
\end{equation*}%
\begin{equation*}
b_{4n}^{-}=C_{41}^{-}u_{1n}^{-}+C_{43}^{-}u_{3n}^{-}+C_{44}^{-}u_{4n}^{%
\shortmid -}+C_{45}^{-}u_{5n}^{\shortmid -};g_{rn\nu }^{-}=f_{n\nu
}^{e+}+f_{n\nu }^{a+}-f_{n\nu }^{o+};
\end{equation*}%
\begin{equation*}
b_{5n}^{-}=C_{52}^{-}u_{2n}^{-}+C_{53}^{-}u_{3n}^{-}+C_{54}^{-}u_{4n}^{%
\shortmid -}+C_{55}^{-}u_{5n}^{\shortmid -};C_{11}^{+}=-\sigma _{1}\frac{%
2\varepsilon _{\nu }\left( -1\right) ^{\nu }}{l};
\end{equation*}%
\begin{equation*}
C_{12}^{+}=-a^{2}\sigma _{1}\frac{2\varepsilon _{\nu }\left( -1\right) ^{\nu
}}{l};C_{14}^{+}=\sigma _{2}n\frac{2\varepsilon _{\nu }\left( -1\right)
^{\nu }}{l};C_{21}^{+}=C_{12}^{+};
\end{equation*}%
\begin{equation*}
C_{22}^{+}=-a^{2}\sigma _{2}\frac{2\varepsilon _{\nu }\left( -1\right) ^{\nu
}}{l};C_{25}^{+}=a^{2}\sigma _{2}n\frac{2\varepsilon _{\nu }\left( -1\right)
^{\nu }}{l};C_{33}^{+}=\varkappa _{1}^{2}\frac{2\varepsilon _{\nu }\left(
-1\right) ^{\nu }}{l};
\end{equation*}%
\begin{equation*}
C_{34}^{+}=-\nu _{01}\frac{2\varepsilon _{\nu }\left( -1\right) ^{\nu }}{l}%
;C_{35}^{+}=C_{33}^{+};C_{44}^{+}=\beta _{\nu }\frac{2\varepsilon _{\nu
}\left( -1\right) ^{\nu }}{l};C_{45}^{+}=a^{2}\beta _{\nu }\frac{%
2\varepsilon _{\nu }\left( -1\right) ^{\nu }}{l};
\end{equation*}%
\begin{equation*}
C_{54}^{+}=C_{55}^{+}+C_{45}^{+};C_{11}^{-}=\sigma _{1}\frac{2\left(
-1\right) ^{\nu }\beta _{\nu }}{l};C_{12}^{-}=a^{2}\frac{2\left( -1\right)
^{\nu }\beta _{\nu }}{l};C_{21}^{-}=C_{12}^{-};
\end{equation*}%
\begin{equation*}
C_{33}^{-}=-\varkappa _{1}^{2}\frac{2\left( -1\right) ^{\nu }\beta _{\nu }}{l%
};C_{33}^{-}=-\sigma _{2}h\frac{2\left( -1\right) ^{\nu }\beta _{\nu }}{l}%
;C_{53}^{-}=C_{33}^{-};C_{54}^{-}=C_{55}^{-}=C_{45}^{-};
\end{equation*}%
\begin{equation*}
C_{52}^{-}=a^{2}C_{41}^{-};C_{45}^{\pm }=\pm C_{12}^{\pm };
\end{equation*}%
From (\ref{43}) we define $U_{jn\nu }^{\pm }$ as:
\begin{equation*}
u_{nj\nu }^{+}=L_{1n\nu }^{j+}u_{1n}^{\shortmid +}+L_{2n\nu
}^{j+}u_{2n}^{\shortmid +}+L_{3n\nu }^{j+}u_{3n}^{\shortmid +}+L_{4n\nu
}^{j+}u_{4n}^{+}+L_{5n\nu }^{j+}u_{5n}^{+}+L_{6n\nu }^{j+}g_{nr\nu }^{+};
\end{equation*}%
\begin{equation}
u_{nj\nu }^{-}=L_{1n\nu }^{j-}u_{1n}^{-}+L_{2n\nu }^{j-}u_{2n}^{-}+L_{3n\nu
}^{j-}u_{3n}^{-}+L_{4n\nu }^{j-}u_{4n}^{\shortmid -}+L_{5n\nu
}^{j-}u_{5n}^{\shortmid -}+L_{6n\nu }^{j-}g_{nr\nu }^{-};j=\overline{1,5};
\label{44}
\end{equation}%
where
\begin{equation*}
L_{1n\nu }^{j+}=\frac{1}{\Delta _{5}^{+}}\left(
C_{11}^{+}A_{1j}^{+}+C_{21}^{+}A_{2j}^{+}\right) ;L_{2n\nu }^{j+}=\frac{1}{%
\Delta _{5}^{+}}\left( C_{12}^{+}A_{1j}^{+}+C_{22}^{+}A_{2j}^{+}\right) ;
\end{equation*}%
\begin{equation*}
L_{3n\nu }^{j+}=\frac{1}{\Delta _{5}^{+}}\left( C_{33}^{+}A_{3j}^{+}\right)
;L_{6n\nu }^{j+}=\frac{A_{3j}^{+}}{\Delta _{5}^{+}};
\end{equation*}%
\begin{equation*}
L_{4n\nu }^{j+}=\frac{1}{\Delta _{5}^{+}}\left( C_{\
44}^{+}A_{4j}^{+}+C_{54}^{+}A_{5j}^{+}+C_{\
14}^{+}A_{1j}^{+}+C_{34}^{+}A_{3j}^{+}\right) ;
\end{equation*}%
\begin{equation*}
L_{5n\nu }^{j+}=\frac{1}{\Delta _{5}^{+}}\left( C_{\
45}^{+}A_{4j}^{+}+C_{55}^{+}A_{5j}^{+}+C_{\
25}^{+}A_{2j}^{+}+C_{35}^{+}A_{3j}^{+}\right) ;
\end{equation*}%
\begin{equation*}
L_{1n\nu }^{j-}=\frac{1}{\Delta _{5}^{-}}\left( C_{\
11}^{-}A_{1j}^{-}+C_{21}^{-}A_{2j}^{-}+C_{\ 41}^{-}A_{2j}^{-}\right) ;
\end{equation*}%
\begin{equation*}
L_{2n\nu }^{j-}=\frac{1}{\Delta _{5}^{-}}\left( C_{\
12}^{-}A_{1j}^{-}+C_{22}^{-}A_{2j}^{-}+C_{\ 52}^{-}A_{5j}^{-}\right) ;
\end{equation*}%
\begin{equation*}
L_{3n\nu }^{j-}=\frac{1}{\Delta _{5}^{-}}\left( C_{\
33}^{-}A_{3j}^{-}+C_{43}^{-}A_{4j}^{-}+C_{\ 53}^{-}A_{5j}^{-}\right) ;
\end{equation*}%
\begin{equation*}
L_{4n\nu }^{j-}=\frac{1}{\Delta _{5}^{-}}\left( C_{\
44}^{-}A_{4j}^{-}+C_{54}^{-}A_{5j}^{-}\right) ;L_{5n\nu }^{j-}=\frac{1}{%
\Delta _{5}^{-}}\left( C_{\ 45}^{-}A_{4j}^{-}+C_{55}^{-}A_{5j}^{-}\right) ;
\end{equation*}%
\begin{equation*}
L_{6n\nu }^{j-}=\frac{A_{3j}^{-}}{\Delta _{5}^{-}};\Delta _{5}^{\pm }=\det
\left\vert a_{ij}^{\pm }\right\vert ;i,j=\left( 1,2,3,4,5\right) ;
\end{equation*}%
$A_{i,j}^{\pm }$ are algebraic adjuncts of elements $a_{ij}^{\pm }.$

To improve the convergence of the obtained solution (\ref{44}) we use the
asymptotic properties of expansion coefficients of Fourier series of the
desired functions. Basing on (\ref{24}), (\ref{26}) we could define the
following relations to the even functions $f\left( z\right) $ and odd
functions $\varphi \left( z\right) $ and its derivatives represented as
Fourier series on $\left[ -l,l\right] ,$

a) for even function $f\left( z\right) $%
\begin{equation}
f\left( z\right) =\sum_{\nu =0}^{N}f_{\nu }\cos \beta _{\nu }z+\sum_{\nu
=N+1}^{\infty }\frac{2\varepsilon _{\nu }\left( -1\right) ^{\nu }}{l}\left[
\frac{f^{\shortmid }\left( l\right) }{\beta _{\nu }^{2}}-\frac{f^{\shortmid
\shortmid \shortmid }\left( l\right) }{\beta _{\nu }^{4}}+\frac{f^{\text{v}%
}\left( l\right) }{\beta _{\nu }^{6}}\right] \cos \beta _{\nu }z;  \label{45}
\end{equation}%
\begin{equation*}
\frac{\partial ^{2}f\left( z\right) }{\partial z^{2}}=\sum_{\nu =0}^{N}\left[
-\beta _{\nu }f_{\nu }+\frac{2\varepsilon _{\nu }\left( -1\right) ^{\nu }}{l}%
f^{\shortmid }\left( l\right) \right] \cos \beta _{\nu }z+\sum_{\nu
=N+1}^{\infty }\frac{2\varepsilon _{\nu }\left( -1\right) ^{\nu }}{l}\left[
\frac{f^{\shortmid \shortmid \shortmid }\left( l\right) }{\beta _{\nu }^{4}}-%
\frac{f^{\text{v}}\left( l\right) }{\beta _{\nu }^{6}}\right] \cos \beta
_{\nu }z
\end{equation*}%
\begin{equation*}
\frac{\partial ^{4}f\left( z\right) }{\partial z^{4}}=\sum_{\nu =0}^{N}\left[
\beta _{\nu }^{4}f_{\nu }+\frac{2\varepsilon _{\nu }\left( -1\right) ^{\nu }%
}{l}\left( \beta _{\nu }^{2}f^{\shortmid }\left( l\right) -f^{\shortmid
\shortmid \shortmid }\left( l\right) \right) \right] \cos \beta _{\nu }z+
\end{equation*}%
\begin{equation*}
+\sum_{\nu =N+1}^{\infty }\frac{2\varepsilon _{\nu }\left( -1\right) ^{\nu }%
}{l}\left[ \frac{f^{\text{v}}\left( l\right) }{\beta _{\nu }^{2}}-\frac{f^{%
\text{v}\shortmid \shortmid }\left( l\right) }{\beta _{\nu }^{6}}\right]
\cos \beta _{\nu }z;f^{\shortmid \shortmid \shortmid }\left( l\right) =\frac{%
\partial ^{\shortmid }f\left( z\right) }{\partial z}\mid _{z=l};
\end{equation*}%
\begin{equation*}
f^{\shortmid \shortmid \shortmid }\left( l\right) =\frac{\partial
^{3}f\left( z\right) }{\partial z^{3}}\mid _{z=l};f^{\text{v}}\left(
l\right) =\frac{\partial ^{5}f\left( z\right) }{\partial z^{5}}\mid
_{z=l};f^{\text{v}\shortmid \shortmid }\left( l\right) =\frac{\partial
^{7}f\left( z\right) }{\partial z^{7}}\mid _{z=l};
\end{equation*}%
etc.

b) for odd function
\begin{equation*}
\varphi \left( z\right) =\sum_{\nu =0}^{N}\varphi _{\nu }\sin \beta _{\nu }z;
\end{equation*}%
\begin{equation}
\frac{\partial \varphi \left( z\right) }{\partial z}=\sum_{\nu =1}^{N}\left[
\beta _{\nu }\varphi _{\nu }+\frac{2\left( -1\right) ^{\nu }}{l}\varphi
\left( l\right) \right] \cos \beta _{\nu }z+\sum_{\nu =N+1}^{\infty }\frac{%
2\left( -1\right) ^{\nu }}{l}\left[ \frac{\varphi ^{\shortmid \shortmid
}\left( l\right) }{\beta _{\nu }}-\frac{\varphi ^{\shortmid \text{v}}\left(
l\right) }{\beta _{\nu }^{4}}\right] \cos \beta _{\nu }z  \label{46}
\end{equation}%
\begin{equation*}
\frac{\partial ^{3}\varphi \left( z\right) }{\partial z^{3}}=\sum_{\nu
=1}^{N}\left[ -\beta _{\nu }^{3}\varphi _{\nu }-\frac{2\left( -1\right)
^{\nu }}{l}\left( \beta _{\nu }^{2}\varphi \left( l\right) -\varphi
^{\shortmid \shortmid }\left( l\right) \right) \right] \cos \beta _{\nu }z+
\end{equation*}%
\begin{equation*}
+\sum_{\nu =N+1}^{\infty }\frac{2\left( -1\right) ^{\nu }}{l}\left[ \frac{%
\varphi ^{\shortmid \text{v}}\left( l\right) }{\beta _{\nu }^{2}}-\frac{%
\varphi ^{\text{v}\shortmid }\left( l\right) }{\beta _{\nu }^{4}}\right]
\cos \beta _{\nu }z;\frac{\partial ^{2}\varphi \left( z\right) }{\partial
z^{2}}\mid _{z=l}=\varphi ^{\shortmid \shortmid }\left( l\right) ;
\end{equation*}%
\begin{equation*}
\frac{\partial ^{5}\varphi \left( z\right) }{\partial z^{5}}=\sum_{\nu
=1}^{N}\left[ \beta _{\nu }^{5}\varphi _{\nu }+\frac{2\left( -1\right) ^{\nu
}}{l}\left( \beta _{\nu }^{4}\varphi \left( l\right) -\beta _{\nu }\varphi
^{\shortmid \shortmid }\left( l\right) \right) \right] \cos \beta _{\nu }z+
\end{equation*}%
\begin{equation*}
+\sum_{\nu =N+1}^{\infty }\frac{2\left( -1\right) ^{\nu }}{l}\left[ \frac{%
\varphi ^{\shortmid \text{v}}\left( l\right) }{\beta _{\nu }^{2}}-\frac{%
\varphi ^{\text{v}\shortmid }\left( l\right) }{\beta _{\nu }^{4}}\right]
\cos \beta _{\nu }z;\frac{\partial ^{4}\varphi \left( z\right) }{\partial
z^{4}}\mid _{z=l}=\varphi ^{\shortmid \text{v}}\left( l\right) ;\frac{%
\partial ^{6}\varphi \left( z\right) }{\partial z^{6}}\mid _{z=l}=\varphi ^{%
\text{v}\shortmid }\left( l\right) .
\end{equation*}

Let us introduce the symbolism
\begin{equation*}
X_{1}^{+}=u_{1n}^{\shortmid +};X_{2}^{+}=u_{2n}^{\shortmid
+};X_{3}^{+}=u_{3n}^{\shortmid +};X_{4}^{+}=u_{4n}^{+};X_{5}^{+}=u_{5n}^{+};
\end{equation*}%
\begin{equation}
X_{1}^{-}=u_{1n}^{-};X_{2}^{-}=u_{2n}^{-};X_{3}^{-}=u_{3n}^{-};X_{4}^{-}=u_{4n}^{\shortmid -};X_{5}^{-}=u_{5n}^{\shortmid -};
\label{47}
\end{equation}%
\begin{equation*}
X_{6}^{\pm }=A_{1n}^{\pm };X_{7}^{\pm }=A_{2n}^{\pm };X_{8}^{\pm }=\Phi
_{1n}^{\pm };X_{9}^{\pm }=\varphi _{1n}^{\pm };X_{10}^{\pm }=\Psi _{1n}^{\pm
}.
\end{equation*}

Substituting obtained solutions of differential equations (\ref{2})-(\ref{8}%
) into the boundary conditions (\ref{9a})-(\ref{12}) we get the system of
equations relative to unknown constants (\ref{47}) with asymptotic (\ref{45}%
), (\ref{46}) for $U_{jn\nu }^{\pm }$ taken into account:
\begin{equation}
\left( a^{\pm }\right) _{ij}\left\{ X^{\pm }\right\} _{i}=\left\{ F^{\pm
}\right\} _{i};  \label{48}
\end{equation}%
where
\begin{equation*}
a_{11}^{+}=\sum_{\nu =0}^{N}L_{1n\nu }^{3+}\left( -1\right) ^{\nu
};a_{12}^{+}=\sum_{\nu =0}^{N}L_{2n\nu }^{3+}\left( -1\right) ^{\nu
};a_{13}^{+}=\sum_{\nu =0}^{N}L_{3n\nu }^{3+}\left( -1\right) ^{\nu
}+B_{1}^{\ast };a_{14}^{+}=\sum_{\nu =0}^{N}L_{4n\nu }^{3+}\left( -1\right) ;
\end{equation*}%
\begin{equation*}
a_{15}^{+}=\sum_{\nu =0}^{N}L_{5n\nu }^{3+}\left( -1\right) ^{\nu
};a_{25}^{+}=1;a_{19}^{\pm }=-k_{1}J_{n}^{\shortmid }\left( k_{1}\right)
;a_{110}^{\pm }=-nJ_{n}\left( k_{2}\right) ;B_{1}^{\ast }=\frac{2l}{\pi }%
\sum_{\nu =N+1}^{\infty }\frac{1}{\nu ^{2}};
\end{equation*}%
\begin{equation*}
a_{31}^{+}=\sum_{\nu =0}^{N}L_{1n\nu }^{1+}\left( -1\right) ^{\nu
}+B_{1}^{\ast };a_{32}^{+}=\sum_{\nu =0}^{N}L_{2n\nu }^{1+}\left( -1\right)
^{\nu };a_{33}^{+}=\sum_{\nu =0}^{N}L_{3n\nu }^{1+}\left( -1\right) ^{\nu
};a_{34}^{+}=\sum_{\nu =0}^{N}L_{4n\nu }^{1+}\left( -1\right) ^{\nu };
\end{equation*}%
\begin{equation*}
a_{26}^{\pm }=\frac{J_{n}\left( \gamma _{1}\right) }{\gamma _{1}}%
;a_{27}^{\pm }=\frac{J_{n}\left( \gamma _{2}\right) }{\gamma _{2}}%
;a_{310}^{\pm }=k_{2}J_{n}^{\shortmid }\left( k_{2}\right) ;a_{39}^{\pm
}=-nJ_{n}\left( k_{1}\right) ;
\end{equation*}%
\begin{equation*}
a_{35}^{+}=\sum_{\nu =0}^{N}L_{5n\nu }^{1+}\left( -1\right) ^{\nu
};a_{41}^{+}=\sum_{\nu =0}^{N}L_{1n\nu }^{2+}\left( -1\right) ^{\nu
};a_{42}^{+}=\sum_{\nu =0}^{N}L_{2n\nu }^{2+}\left( -1\right) ^{\nu
}+B_{1}^{\ast };a_{43}^{+}=\sum_{\nu =0}^{N}L_{3n\nu }^{2+}\left( -1\right)
^{\nu };
\end{equation*}%
\begin{equation*}
a_{44}^{+}=\sum_{\nu =0}^{N}L_{4n\nu }^{2+}\left( -1\right) ^{\nu
};a_{45}^{+}=\sum_{\nu =0}^{N}L_{3n\nu }^{2+}\left( -1\right) ^{\nu
};a_{54}^{+}=1;a_{49}^{\pm }=\frac{n}{R_{0}}\left( k_{1}J_{n}^{\shortmid
}\left( k_{1}\right) -J_{n}\left( k_{1}\right) \right) ;
\end{equation*}%
\begin{equation*}
a_{410}^{\pm }=\frac{nJ_{n}\left( k_{2}\right) }{R_{0}}-\frac{%
k_{2}J_{n}^{\shortmid }\left( k_{2}\right) }{R_{0}};a_{56}^{\pm }=\left( -%
\frac{1}{\gamma _{1}}+S\gamma _{1}\right) J_{n}^{\shortmid }\left( \gamma
_{1}\right) ;a_{57}^{\pm }=\left( -\frac{1}{\gamma _{2}}+S\gamma _{2}\right)
J_{n}^{\shortmid }\left( \gamma _{2}\right) ;
\end{equation*}%
\begin{equation*}
a_{61}^{+}=B_{1}\left\{ \sum_{\nu =1}^{N}L_{1n\nu }^{1+}\left( -1\right)
^{\nu }\beta _{\nu }+\frac{\nu _{01}}{R_{0}}n\left[ \sum_{\nu
=0}^{N}L_{1n\nu }^{1+}\left( -1\right) ^{\nu }+B_{1}^{\ast }\right] +\frac{%
\nu _{01}}{R_{0}}\sum_{\nu =0}^{N}L_{1n\nu }^{3+}\left( -1\right) ^{\nu
}\right\}
\end{equation*}%
\begin{equation*}
a_{58}^{\pm }=-k_{3}J_{n}^{\shortmid }\left( k_{3}\right)
;a_{62}^{+}=B_{1}\left\{ \sum_{\nu =1}^{N}L_{2n\nu }^{1+}\left( -1\right)
^{\nu }\beta _{\nu }+\frac{\nu _{01}}{R_{0}}n\sum_{\nu =0}^{N}L_{1n\nu
}^{1+}\left( -1\right) ^{\nu }+\frac{\nu _{01}}{R_{0}}\sum_{\nu
=0}^{N}L_{1n\nu }^{3+}\left( -1\right) ^{\nu }\right\}
\end{equation*}%
\begin{equation*}
a_{63}^{+}=B_{1}\left\{ \sum_{\nu =1}^{N}L_{3n\nu }^{5+}\left( -1\right)
^{\nu }\beta _{\nu }+\frac{\nu _{01}}{R_{0}}n\sum_{\nu =0}^{N}L_{3n\nu
}^{1+}\left( -1\right) ^{\nu }+\frac{\nu _{01}}{R_{0}}\left[ \sum_{\nu
=0}^{N}L_{1n\nu }^{1+}\left( -1\right) ^{\nu }+B_{1}^{\ast }\right] \right\}
\end{equation*}%
\begin{equation*}
a_{64}^{+}=B_{1}\left\{ \sum_{\nu =1}^{N}L_{4n\nu }^{5+}\left( -1\right)
^{\nu }\beta _{\nu }+\frac{\nu _{01}}{R_{0}}n\sum_{\nu =0}^{N}L_{4n\nu
}^{1+}\left( -1\right) ^{\nu }+\frac{\nu _{01}}{R_{0}}\sum_{\nu
=0}^{N}L_{4n\nu }^{3+}\left( -1\right) ^{\nu }\right\}
\end{equation*}%
\begin{equation*}
a_{65}^{+}=B_{1}\left\{ \sum_{\nu =1}^{N}\left[ L_{5n\nu }^{5+}\left(
-1\right) ^{\nu }\beta _{\nu }+\frac{2N}{l}\right] +\frac{\nu _{01}}{R_{0}}%
n\sum_{\nu =0}^{N}L_{5n\nu }^{1+}\left( -1\right) ^{\nu }+\frac{\nu _{01}}{%
R_{0}}\sum_{\nu =0}^{N}L_{5n\nu }^{3+}\left( -1\right) ^{\nu }\right\}
\end{equation*}%
\begin{equation*}
a_{66}^{\pm }=-\Lambda ^{\shortmid }S\gamma _{1}J_{n}^{\shortmid }\left(
\gamma _{1}\right) ;a_{67}^{\pm }=-\Lambda ^{\shortmid }S\gamma
_{2}J_{n}^{\shortmid }\left( \gamma _{2}\right) ;a_{68}^{\pm }=-nJ_{n}\left(
k_{3}\right) ;a_{74}^{+}=a_{75}^{+}=\Lambda ^{\shortmid };
\end{equation*}%
\begin{equation*}
a_{79}^{\pm }=-\frac{E_{2}h_{2}}{\left( 1-\nu _{02}^{2}\right) R_{0}}\left[
k_{1}^{2}J_{n}^{\shortmid \shortmid }\left( k_{1}\right) +\nu
_{02}k_{1}J_{n}^{\shortmid }\left( k_{1}\right) -\nu _{02}n^{2}J_{n}\left(
k_{1}\right) \right] ;
\end{equation*}%
\begin{equation*}
a_{710}^{\pm }=-\frac{E_{2}h_{2}}{\left( 1-\nu _{02}^{2}\right) R_{0}}\left[
nk_{2}J_{n}^{\shortmid }\left( k_{2}\right) +\nu _{02}nJ_{n}\left(
k_{2}\right) -\nu _{02}nJ_{n}^{\shortmid }\left( k_{2}\right) \right] ;
\end{equation*}%
\begin{equation*}
a_{81}^{+}=B_{1}\frac{1-\nu _{01}}{2};a_{85}^{+}=B_{1}\frac{1-\nu _{01}}{2}%
\frac{n}{R_{0}};
\end{equation*}%
\begin{equation*}
a_{89}^{\pm }=-\frac{E_{2}h_{2}}{\left( 1-\nu _{02}^{2}\right) R_{0}}\left[
-nk_{1}J_{n}^{\shortmid }\left( k_{1}\right) -nk_{1}J_{n}^{\shortmid }\left(
k_{1}\right) +nJ_{n}\left( k_{1}\right) \right] ;
\end{equation*}%
\begin{equation*}
a_{810}^{\pm }=-\frac{E_{2}h_{2}}{\left( 1-\nu _{02}^{2}\right) R_{0}}\left[
-n^{2}J_{n}\left( k_{2}\right) -k_{2}^{2}J_{n}^{\shortmid \shortmid }\left(
k_{2}\right) +k_{2}J_{n}^{\shortmid }\left( k_{1}\right) \right] ;
\end{equation*}%
\begin{equation*}
a_{92}^{+}=D_{1}\left\{ \sum_{\nu =1}^{N}L_{2n\nu }^{4+}\left( -1\right)
^{\nu }\beta _{\nu }+\frac{n\nu _{01}}{R_{0}}\left[ \sum_{\nu
=0}^{N}L_{4n\nu }^{2+}\left( -1\right) ^{\nu }+B_{1}^{\ast }\right] \right\}
;
\end{equation*}%
\begin{equation*}
a_{91}^{+}=D_{1}\left\{ \sum_{\nu =1}^{N}L_{1n\nu }^{4+}\left( -1\right)
^{\nu }\beta _{\nu }+\frac{n\nu _{01}}{R_{0}}\sum_{\nu =0}^{N}L_{1n\nu
}^{2+}\left( -1\right) ^{\nu }\right\} ;
\end{equation*}%
\begin{equation*}
a_{93}^{+}=D_{1}\left\{ \sum_{\nu =1}^{N}L_{3n\nu }^{4+}\left( -1\right)
^{\nu }\beta _{\nu }+\frac{n\nu _{01}}{R_{0}}\sum_{\nu =0}^{N}L_{3n\nu
}^{2+}\left( -1\right) ^{\nu }\right\} ;
\end{equation*}%
\begin{equation*}
a_{94}^{+}=D_{1}\left\{ \sum_{\nu =1}^{N}\left[ L_{4n\nu }^{4+}\left(
-1\right) ^{\nu }\beta _{\nu }+\frac{2N}{l}\right] +\frac{\nu _{01}}{R_{0}}%
n\sum_{\nu =0}^{N}L_{4n\nu }^{2+}\left( -1\right) ^{\nu }\right\} ;
\end{equation*}%
\begin{equation*}
a_{95}^{+}=D_{1}\left\{ \sum_{\nu =1}^{N}L_{5n\nu }^{4+}\left( -1\right)
^{\nu }\beta _{\nu }+\frac{n\nu _{01}}{R_{0}}\sum_{\nu =0}^{N}L_{5n\nu
}^{2+}\left( -1\right) ^{\nu }\right\}
\end{equation*}%
\begin{equation*}
a_{96}^{\pm }=D_{1}\left[ \left( \gamma _{1}^{2}S-1\right) J_{n}^{\shortmid
\shortmid }\left( \gamma _{1}\right) +\nu _{02}\left( S\gamma _{1}-\frac{1}{%
\gamma _{1}}\right) J_{n}^{\shortmid }\left( \gamma _{1}\right) +\nu
_{02}n^{2}\left( \frac{1-\gamma _{1}^{2}}{\gamma _{1}^{2}}\right) \right] ;
\end{equation*}%
\begin{equation*}
a_{97}^{\pm }=D_{1}\left[ \left( \gamma _{1}^{2}S-1\right) J_{n}^{\shortmid
\shortmid }\left( \gamma _{2}\right) +\nu _{02}\left( S\gamma _{2}-\frac{1}{%
\gamma _{2}}\right) J_{n}^{\shortmid }\left( \gamma _{2}\right) +\nu
_{02}n^{2}\left( \frac{1-\gamma _{2}^{2}}{\gamma _{2}^{2}}\right) \right] ;
\end{equation*}%
\begin{equation*}
a_{98}^{+}=-D_{2}\left( \frac{1-\nu _{02}}{2}\right) k_{3}J_{n}\left(
k_{3}\right) ;a_{101}^{+}=D_{2}\left( \frac{1-\nu _{02}}{2R_{0}}\right)
;a_{102}^{+}=D_{1}\left( \frac{1-\nu _{02}}{2}\right) ;
\end{equation*}%
\begin{equation*}
a_{104}^{+}=-D_{1}\left( \frac{1-\nu _{02}}{2}\right) \frac{n}{R_{0}}%
;a_{106}^{+}=-D_{2}n\left( 1-\nu _{02}\right) \left( S\gamma _{1}-\frac{1}{%
\gamma _{1}}\right) J_{n}^{\shortmid }\left( \gamma _{1}\right) ;
\end{equation*}%
\begin{equation*}
a_{107}^{+}=-D_{2}n\left( 1-\nu _{02}\right) \left( S\gamma _{2}-\frac{1}{%
\gamma _{2}}\right) J_{n}^{\shortmid }\left( \gamma _{2}\right)
;a_{13}^{-}=1;
\end{equation*}%
\begin{equation*}
a_{98}^{+}=D_{2}\left( \frac{1-\nu _{02}}{2}\right) \left[ 2n^{2}J_{n}\left(
k_{3}\right) +2k_{3}J_{n}^{\shortmid }\left( k_{3}\right)
-k_{3}^{2}J_{n}\left( k_{3}\right) \right] ;
\end{equation*}%
\begin{equation*}
a_{21}^{-}=\sum_{\nu =1}^{N}L_{1n\nu }^{5-}\left( -1\right) ^{\nu
};a_{22}^{-}=\sum_{\nu =1}^{N}L_{2n\nu }^{5-}\left( -1\right) ^{\nu
};a_{23}^{-}=\sum_{\nu =1}^{N}L_{3n\nu }^{5-}\left( -1\right) ^{\nu
};a_{24}^{-}=\sum_{\nu =1}^{N}L_{4n\nu }^{5-}\left( -1\right) ^{\nu };
\end{equation*}%
\begin{equation*}
a_{25}^{-}=\sum_{\nu =1}^{N}L_{5n\nu }^{5-}\left( -1\right) ^{\nu
}+B_{1}^{\ast };a_{51}^{-}=\sum_{\nu =1}^{N}L_{1n\nu }^{4-}\left( -1\right)
^{\nu };a_{52}^{-}=\sum_{\nu =1}^{N}L_{2n\nu }^{4-}\left( -1\right) ^{\nu
};a_{31}^{-}=a_{42=}^{-}1;
\end{equation*}%
\begin{equation*}
a_{51}^{-}=\sum_{\nu =1}^{N}L_{1n\nu }^{4-}\left( -1\right) ^{\nu
};a_{52}^{-}=\sum_{\nu =1}^{N}L_{2n\nu }^{4-}\left( -1\right) ^{\nu
};a_{53}^{-}=\sum_{\nu =1}^{N}L_{3n\nu }^{4-}\left( -1\right) ^{\nu
};a_{54}^{-}=\sum_{\nu =1}^{N}L_{4n\nu }^{4-}\left( -1\right) ^{\nu
}+B_{1}^{\ast };
\end{equation*}%
\begin{equation*}
a_{55}^{-}=\sum_{\nu =1}^{N}L_{5n\nu }^{4-}\left( -1\right) ^{\nu
};a_{61}^{-}=B_{1}\frac{\nu _{01}n}{R_{0}};a_{63}^{-}=B_{1}\frac{\nu _{01}}{%
R_{0}};a_{65}^{-}=B_{1};
\end{equation*}%
\begin{equation*}
a_{71}^{-}=\Lambda _{1}\left\{ \sum_{\nu =1}^{N}L_{1n\nu }^{3-}\left(
-1\right) ^{\nu }\beta _{\nu }+\sum_{\nu =0}^{N}L_{1n\nu }^{4-}\left(
-1\right) ^{\nu }\right\} ;
\end{equation*}%
\begin{equation*}
a_{72}^{-}=\Lambda _{1}\left\{ \sum_{\nu =1}^{N}L_{2n\nu }^{3-}\left(
-1\right) ^{\nu }\beta _{\nu }+\frac{n\nu _{01}}{R_{0}}\sum_{\nu
=0}^{N}L_{2n\nu }^{4-}\left( -1\right) ^{\nu }\right\} ;
\end{equation*}%
\begin{equation*}
a_{73}^{-}=\Lambda _{1}\left\{ \sum_{\nu =1}^{N}L_{3n\nu }^{3-}\left(
-1\right) ^{\nu }\beta _{\nu }+\frac{2N}{l}+\sum_{\nu =0}^{N}L_{3n\nu
}^{4-}\left( -1\right) ^{\nu }\right\}
\end{equation*}%
\begin{equation*}
a_{74}^{-}=\Lambda _{1}\left\{ \sum_{\nu =1}^{N}L_{4n\nu }^{3-}\left(
-1\right) ^{\nu }\beta _{\nu }+\left[ \sum_{\nu =0}^{N}L_{3n\nu }^{4-}\left(
-1\right) ^{\nu }+B_{1}^{\ast }\right] \right\}
\end{equation*}%
\begin{equation*}
a_{75}^{-}=\Lambda _{1}\left\{ \sum_{\nu =1}^{N}L_{5n\nu }^{3-}\left(
-1\right) ^{\nu }\beta _{\nu }+\sum_{\nu =0}^{N}L_{5n\nu }^{4-}\left(
-1\right) ^{\nu }\right\} ;
\end{equation*}%
\begin{equation*}
a_{76}^{-}=-\Lambda _{1}\left\{ \sum_{\nu =1}^{N}\left[ L_{5n\nu
}^{3-}\left( -1\right) ^{\nu }\beta _{\nu }+L_{5n\nu }^{4-}\left( -1\right)
^{\nu }\right] g_{rn\nu }^{-}\right\}
\end{equation*}%
\begin{equation*}
a_{81}^{-}=B_{1}\frac{1-\nu _{02}}{2}\left\{ -\frac{n}{R_{0}}\sum_{\nu
=1}^{N}L_{1n\nu }^{5-}\left( -1\right) ^{\nu }+\left[ \sum_{\nu
=0}^{N}L_{1n\nu }^{1-}\left( -1\right) ^{\nu }\beta _{\nu }+\frac{2N}{l}%
\right] \right\} ;
\end{equation*}%
\begin{equation*}
a_{82}^{-}=B_{1}\frac{1-\nu _{02}}{2}\left\{ -\frac{n}{R_{0}}\sum_{\nu
=1}^{N}L_{2n\nu }^{5-}\left( -1\right) ^{\nu }+\sum_{\nu =0}^{N}L_{2n\nu
}^{1-}\left( -1\right) ^{\nu }\beta _{\nu }\right\} ;
\end{equation*}%
\begin{equation*}
a_{83}^{-}=B_{1}\frac{1-\nu _{02}}{2}\left\{ -\frac{n}{R_{0}}\sum_{\nu
=1}^{N}L_{3n\nu }^{5-}\left( -1\right) ^{\nu }+\sum_{\nu =0}^{N}L_{3n\nu
}^{1-}\left( -1\right) ^{\nu }\beta _{\nu }\right\} ;
\end{equation*}%
\begin{equation*}
a_{84}^{-}=B_{1}\frac{1-\nu _{02}}{2}\left\{ -\frac{n}{R_{0}}\sum_{\nu
=1}^{N}L_{4n\nu }^{5-}\left( -1\right) ^{\nu }+\sum_{\nu =1}^{N}L_{4n\nu
}^{1-}\left( -1\right) ^{\nu }\beta _{\nu }\right\} ;
\end{equation*}%
\begin{equation*}
a_{85}^{-}=B_{1}\frac{1-\nu _{02}}{2}\left\{ -\frac{n}{R_{0}}\left[
\sum_{\nu =1}^{N}L_{5n\nu }^{5-}\left( -1\right) ^{\nu }+B_{1}^{\ast }\right]
+\sum_{\nu =0}^{N}L_{5n\nu }^{1-}\left( -1\right) ^{\nu }\beta _{\nu
}\right\}
\end{equation*}%
\begin{equation*}
a_{101}^{-}=D_{1}\frac{1-\nu _{02}}{2}\left\{ -\frac{1}{R_{0}}\left[
\sum_{\nu =1}^{N}L_{1n\nu }^{1-}\left( -1\right) ^{\nu }\beta _{\nu }+\frac{%
2N}{l}\right] +\sum_{\nu =0}^{N}L_{1n\nu }^{2-}\left( -1\right) ^{\nu }\beta
_{\nu }-\frac{n}{R_{0}}\sum_{\nu =1}^{N}L_{4n\nu }^{4-}\left( -1\right)
^{\nu }\right\}
\end{equation*}%
\begin{equation*}
a_{102}^{-}=D_{1}\frac{1-\nu _{02}}{2}\left\{
\begin{array}{c}
\left[ \sum_{\nu =1}^{N}L_{2n\nu }^{2-}\left( -1\right) ^{\nu }\beta _{\nu }-%
\frac{2N}{l}\right] - \\
-\frac{n}{R_{0}}\sum_{\nu =0}^{N}L_{4n\nu }^{4-}\left( -1\right) ^{\nu }-%
\frac{1}{R_{0}}\sum_{\nu =1}^{N}L_{4n\nu }^{4-}\left( -1\right) ^{\nu }\beta
_{\nu };%
\end{array}%
\right\}
\end{equation*}%
\begin{equation*}
a_{92}^{-}=D_{1}\frac{\nu _{01}}{R_{0}}n;a_{94}^{-}=D_{1};
\end{equation*}%
\begin{equation*}
a_{103}^{-}=D_{1}\frac{1-\nu _{02}}{2}\left\{ \sum_{\nu =1}^{N}L_{3n\nu
}^{2-}\left( -1\right) ^{\nu }\beta _{\nu }+\frac{1}{R_{0}}\sum_{\nu
=0}^{N}L_{3n\nu }^{1-}\left( -1\right) ^{\nu }\beta _{\nu }-\frac{n}{R_{0}}%
\sum_{\nu =1}^{N}L_{3n\nu }^{4-}\left( -1\right) ^{\nu }\right\} ;
\end{equation*}%
\begin{equation*}
a_{104}^{-}=D_{1}\frac{1-\nu _{02}}{2}\left\{ -\frac{n}{R_{0}}\left[
\sum_{\nu =1}^{N}L_{4n\nu }^{4-}\left( -1\right) ^{\nu }+B_{1}^{\ast }\right]
+\sum_{\nu =1}^{N}L_{4n\nu }^{2-}\left( -1\right) ^{\nu }\beta _{\nu }-\frac{%
1}{R_{0}}\sum_{\nu =1}^{N}L_{4n\nu }^{1-}\left( -1\right) ^{\nu }\beta _{\nu
};\right\}
\end{equation*}%
\begin{equation*}
a_{105}^{-}=D_{1}\frac{1-\nu _{02}}{2}\left\{ \sum_{\nu =1}^{N}L_{5n\nu
}^{2-}\left( -1\right) ^{\nu }\beta _{\nu }+\frac{1}{R_{0}}\sum_{\nu
=0}^{N}L_{5n\nu }^{1-}\left( -1\right) ^{\nu }\beta _{\nu }-\frac{n}{R_{0}}%
\sum_{\nu =1}^{N}L_{5n\nu }^{4-}\left( -1\right) ^{\nu }\right\} ;
\end{equation*}%
\begin{equation*}
F_{i}^{\ast \pm }=\sum_{\nu =0}^{N}\overline{C}_{in\nu }^{\pm }f_{rn\nu
}^{\pm }+\sum_{\nu =0}^{N}\overline{C}_{in\nu }^{\pm }g_{znj}^{\pm
};i=1,2,3......10
\end{equation*}%
\begin{equation*}
\overline{C}_{1n\nu }^{+}=-L_{6n\nu }^{3+}\left( -1\right) ^{\nu };\overline{%
C}_{2nj}^{\pm }=A_{3nj}^{\ast };\overline{C}_{3n\nu }^{+}=-L_{6n\nu
}^{1+}\left( -1\right) ^{\nu };\overline{C}_{4n\nu }^{+}=-L_{6n\nu
}^{2+}\left( -1\right) ^{\nu };
\end{equation*}%
\begin{equation*}
A_{3nj}^{\ast }=\frac{A_{3nj}-\frac{S}{D_{2}}}{-\gamma _{jn}^{2}}J_{n}\left(
\gamma _{jn}\right) ;\overline{C}_{6n\nu }^{+}=-B_{1}\left[ L_{6n\nu }^{5+}+%
\frac{\nu _{01}n}{R_{0}}L_{6n\nu }^{1+}+\frac{\nu _{01}}{R_{0}}L_{6n\nu
}^{3+}\right] ;
\end{equation*}%
\begin{equation*}
\overline{C}_{9n\nu }^{+}=-D_{1}\left[ L_{6n\nu }^{4+}+\frac{\nu _{01}n}{%
R_{0}}L_{6n\nu }^{2+}\right] ;\overline{C}_{2n\nu }^{-}=-L_{6n\nu
}^{5-}\left( -1\right) ^{\nu };\overline{C}_{5n\nu }^{-}=-L_{6n\nu
}^{4-}\left( -1\right) ^{\nu }
\end{equation*}%
\begin{equation*}
\overline{C}_{9n\nu }^{\shortmid +}=D_{2}\left[ \frac{A_{3nj}-\gamma
_{jn}^{2}SA_{3nj}-\frac{S}{D_{2}}}{\gamma _{jn}^{2}}\left( \gamma
_{jn}^{2}-n^{2}+1\right) \right] J_{n}\left( \gamma _{jn}\right) ;
\end{equation*}%
\begin{equation*}
\overline{C}_{7n\nu }^{-}=-\Lambda _{1}\left\{ L_{6n\nu }^{3-}\left(
-1\right) ^{\nu }\beta _{\nu }+\sum_{\nu =0}^{N}L_{6n\nu }^{4-}\left(
-1\right) ^{\nu }\right\} ;
\end{equation*}%
\begin{equation*}
\overline{C}_{8n\nu }^{-}=-B_{1}\frac{1-\nu _{02}}{2}\left\{ -\frac{n}{R_{0}}%
L_{6n\nu }^{5-}\left( -1\right) ^{\nu }+L_{6n\nu }^{1-}\left( -1\right)
^{\nu }\beta _{\nu }\right\} ;
\end{equation*}%
\begin{equation*}
\overline{C}_{10n\nu }^{-}=-D_{1}\frac{1-\nu _{02}}{2}\left\{ -\frac{n}{R_{0}%
}L_{6n\nu }^{4-}\left( -1\right) ^{\nu }+L_{6n\nu }^{2-}\left( -1\right)
^{\nu }\beta _{\nu }+\frac{1}{R_{0}}L_{6n\nu }^{1-}\left( -1\right) ^{\nu
}\beta _{\nu }\right\} ;
\end{equation*}%
Solution (\ref{48}) will be
\begin{equation}
X_{i}=\sum_{\nu =0}^{N}Z_{n\nu j}^{\pm }f_{rn\nu }^{\pm
}+\sum_{j=0}^{N}Z_{n\nu j}^{\shortmid \pm }g_{rnj}^{\pm };\left(
i=1,2,3....10\right)  \label{49}
\end{equation}%
where
\begin{equation*}
Z_{n\nu j}^{\pm }=\sum_{k=1}^{10}\frac{\overline{C}_{kn\nu }^{\pm }\Delta
_{9ki}^{\pm }}{\Delta _{10}^{\pm }};Z_{n\nu j}^{\shortmid \pm
}=\sum_{k=1}^{10}\frac{\overline{C}_{kn\nu }^{\shortmid \pm }\Delta
_{9ki}^{\pm }}{\Delta _{10}^{\pm }}.
\end{equation*}

For $U_{3n\nu }$ and $U_{zn\mu }$ we obtain
\begin{equation}
u_{3n\mu }^{\pm }=L_{1n\mu }^{3\pm }X_{1}^{\pm }+L_{2n\mu }^{3\pm
}X_{2}^{\pm }+L_{3n\mu }^{3\pm }X_{3}^{\pm }+L_{4n\mu }^{3\pm }X_{4}^{\pm
}+L_{5n\mu }^{3\pm }X_{5}^{\pm }+L_{6n\mu }^{3\pm }f_{rn\mu }^{\pm };
\label{50}
\end{equation}%
\begin{equation*}
W_{zn\mu }^{\pm }=L_{7n\mu }^{\pm }X_{6}^{\pm }+L_{8n\mu }^{\pm }X_{7}^{\pm
}+L_{9n\mu }^{\pm }g_{zn\mu }^{\pm };
\end{equation*}

To improve the convergence of the obtained solution we will single-out the
characteristic features in unknown functions. It is known that \cite{16, 18}
approaching to angle circles on surface $r=1$ and $z=l$ expression for the
pressure $P^{e\pm }$ and its derivatives must have the following
singularities

a) for the pressure
\begin{equation}
P_{n}^{e+}\left( z_{0}\right) =A_{n}\sqrt[3]{\left( l^{2}-z_{0}^{2}\right)
^{2}};P_{n}^{e-}\left( z_{0}\right) =B_{n}z_{0}\sqrt[3]{\left(
l^{2}-z_{0}^{2}\right) };P_{n}^{e\pm }\left( r_{0}\right) =D_{n}^{\pm }\sqrt[%
3]{\left( 1-r_{0}\right) ^{2};}  \label{51}
\end{equation}

b) for the derivatives from pressure
\begin{equation*}
\frac{\partial P_{n}^{e+}\left( z_{0}r\right) }{\partial r}\mid _{r=1}=\frac{%
A_{n}^{\ast }}{\sqrt[3]{l^{2}-z_{0}^{2}}};\frac{\partial P_{n}^{e-}\left(
z_{0}r\right) }{\partial r}\mid _{r=1}=\frac{B_{n}^{\ast }z_{0}}{\sqrt[3]{%
l^{2}-z_{0}^{2}}};
\end{equation*}%
\begin{equation}
\frac{\partial P_{n}^{e-}\left( z_{0}r\right) }{\partial z}\mid _{z=l}=\frac{%
D_{n}^{\ast \pm }r^{n}}{\sqrt[3]{1-r_{0}^{2}}}.  \label{52}
\end{equation}

Expanding (\ref{51}), (\ref{52}) into Fourier's, Fourier-Bessel's series
corresponding to (\ref{14}) we obtain \cite{15, 21}%
\begin{equation*}
A_{n}\sqrt[0]{\left( l^{2}-z_{0}^{2}\right) ^{2}}=\sum_{\nu =0}^{\infty
}f_{n\nu }^{s+}\cos \beta _{\nu }z_{0};B_{n}z_{0}\sqrt[3]{\left(
l^{2}-z_{0}^{2}\right) }=\sum_{\nu =0}^{\infty }f_{n\nu }^{s-}\sin \beta
_{\nu }z_{0};
\end{equation*}%
\begin{equation}
D_{n}^{\pm }\sqrt[3]{\left( 1-r_{0}^{2}\right) ^{2}}=\sum_{\nu =0}^{\infty
}g_{nj}^{s-}J_{n}\left( \gamma _{jn}r\right) ;\frac{A_{n}^{\ast }}{\sqrt[3]{%
l^{2}-z_{0}^{2}}}=\sum_{\nu =0}^{\infty }f_{n\nu }^{\ast s+}\cos \beta _{\nu
}z_{0};  \label{53}
\end{equation}%
\begin{equation*}
\frac{B_{n}^{\ast }z_{0}}{\sqrt[3]{l^{2}-z_{0}^{2}}}=\sum_{\nu =0}^{\infty
}f_{n\nu }^{\ast s-}\sin \beta _{\nu }z_{0};\frac{D_{n}^{\ast \pm }r^{n}}{%
\sqrt[3]{1-r_{0}^{2}}}=\sum_{\nu =0}^{\infty }g_{nj}^{\ast s\pm }J_{n}\left(
\gamma _{jn}r\right)
\end{equation*}%
where
\begin{equation*}
f_{n\nu }^{s+}=\frac{A_{n}l^{1/6}\text{$\Gamma $}\left( 5/3\right) \text{$%
\Gamma $}\left( 1/2\right) \text{$\Gamma $}_{7/6}\left( \beta _{\nu
}l\right) 2^{7/6}}{\beta _{\nu }^{7/6}};g_{nj}^{\ast s\pm }=\frac{%
D_{n}^{\ast \pm }\varepsilon _{nj}\text{$\Gamma $}\left( 2/3\right) \text{$%
\Gamma $}_{n+7/6}\left( \gamma _{jn}\right) }{\sqrt[3]{2\gamma _{jn}^{2}}};
\end{equation*}%
\begin{equation*}
g_{nj}^{s\pm }=\frac{D_{n}^{\pm }\varepsilon _{nj}2^{2/3}\text{$\Gamma $}%
\left( 5/3\right) \text{$\Gamma $}_{n+5/3}\left( \gamma _{jn}\right) }{%
\gamma _{jn}^{5/3}};f_{n\nu }^{\ast s+}=\frac{A_{n}\text{$\Gamma $}\left(
2/3\right) \text{$\Gamma $}\left( 1/2\right) J_{1/6}\left( \beta _{\nu
}l\right) 2^{1/6}}{l^{5/6}\beta _{\nu }^{7/6}};
\end{equation*}%
\begin{equation*}
f_{n\nu }^{s-}=-\frac{A_{n}\text{$\Gamma $}\left( 5/3\right) \text{$\Gamma $}%
\left( 1/2\right) J_{13/6}\left( \beta _{\nu }l\right) 2^{7/6}l^{7/6}}{%
l^{5/6}\beta _{\nu }^{7/6}};f_{n\nu }^{\ast s-}=-\frac{A_{n}\text{$\Gamma $}%
\left( 2/3\right) \text{$\Gamma $}\left( 1/2\right) J_{1/6}\left( \beta
_{\nu }l\right) 2^{1/6}\beta _{\nu }l}{\beta _{\nu }^{7/6}}.
\end{equation*}

Unknown coefficients $f_{n\nu }^{e\pm },$ $f_{n\nu }^{\ast e\pm },$ $%
g_{nj}^{e\pm },$ $g_{nj}^{\ast e\pm }$ at large $\nu ,j\left( \nu
,j>N\right) $ behave itself like coefficients of known expansions (\ref{53}%
), i.e. we could write
\begin{equation}
f_{n\nu }^{e\pm }=\frac{f_{nN}^{e\pm }}{f_{nN}^{s\pm }}f_{n\nu }^{s\pm
};f_{n\nu }^{\ast e\pm }=\frac{f_{nN}^{\ast e\pm }}{f_{nN}^{\ast \pm }}%
f_{n\nu }^{\ast s\pm };g_{n\nu }^{e\pm }=\frac{g_{nN}^{e\pm }}{g_{nN}^{s\pm }%
}g_{n\nu }^{s\pm };g_{n\nu }^{\ast e\pm }=\frac{g_{nN}^{\ast e\pm }}{%
g_{nN}^{\ast \pm }}g_{n\nu }^{\ast s\pm }.  \label{54}
\end{equation}%
Thus, if in the unlimited systems of algebraic linear equations (\ref{23})
unknown coefficients are replaced with formulas (\ref{54}) than elements of $%
N$-th columns in matrixes will be:
\begin{equation*}
F_{n\mu N}^{\ast 1+}=\frac{\left( -1\right) ^{\mu }}{f_{nN}^{\ast s+}}\int
\left\{ \frac{\lambda ^{2}J_{n}^{\shortmid }\left( \lambda \right)
J_{n}\left( \lambda \right) }{\left( \varkappa ^{2}+\beta _{\mu }^{2}\right)
}\varkappa \left( 1-e^{-2\varkappa l}\right) \sum_{\nu =N+1}^{\infty }\frac{%
\left( -1\right) ^{\nu }f_{n\nu }^{\ast s+}}{\varkappa ^{2}+\beta _{\nu }^{2}%
}\right\} d\lambda ;
\end{equation*}%
\begin{equation*}
F_{n\mu N}^{1+}=\frac{\left( -1\right) ^{\mu }}{f_{nN}^{s+}}\int \left\{
\frac{\lambda ^{2}J_{n}^{\shortmid }\left( \lambda \right) J_{n}\left(
\lambda \right) }{\left( \varkappa ^{2}+\beta _{\mu }^{2}\right) }\varkappa
\left( 1-e^{-2\varkappa l}\right) \sum_{\nu =N+1}^{\infty }\frac{\left(
-1\right) ^{\nu }f_{n\nu }^{s+}}{\varkappa ^{2}+\beta _{\nu }^{2}}\right\}
d\lambda ;
\end{equation*}%
\begin{equation*}
G_{n\mu N}^{\ast 1+}=\frac{\left( -1\right) ^{\mu }}{g_{nN}^{\ast s+}}\int
\left\{ \frac{\lambda ^{2}J_{n}^{\shortmid }\left( \lambda \right)
J_{n}\left( \lambda \right) }{\left( \varkappa ^{2}+\beta _{\mu }^{2}\right)
}\varkappa \left( 1-e^{-2\varkappa l}\right) \sum_{j=N+1}^{\infty }\frac{%
g_{nj}^{\ast s+}J_{n}\left( \gamma _{jn}\right) }{\gamma _{jn}^{2}-\lambda
^{2}}\right\} d\lambda ;
\end{equation*}%
\begin{equation*}
G_{n\mu N}^{1+}=\frac{\left( -1\right) ^{\mu }}{g_{nN}^{s+}}\int \left\{
\frac{\lambda ^{2}J_{n}^{\shortmid }\left( \lambda \right) J_{n}\left(
\lambda \right) }{\left( \varkappa ^{2}+\beta _{\mu }^{2}\right) }\varkappa
\left( 1-e^{-2\varkappa l}\right) \sum_{j=N+1}^{\infty }\frac{%
g_{nj}^{s+}J_{n}\left( \gamma _{jn}\right) }{\gamma _{jn}^{2}-\lambda ^{2}}%
\right\} d\lambda ;
\end{equation*}%
\begin{equation*}
F_{n\mu N}^{\ast 2+}=\frac{J_{n}\left( \gamma _{jn}\right) }{f_{nN}^{\ast s+}%
}\int \left\{ \frac{\lambda ^{2}J_{n}^{\shortmid }\left( \lambda \right)
J_{n}\left( \lambda \right) }{\gamma _{jn}^{2}-\lambda ^{2}}\left(
1-e^{-2\varkappa l}\right) \sum_{\nu =N+1}^{\infty }\frac{\left( -1\right)
^{\nu }f_{n\nu }^{\ast s+}}{\varkappa ^{2}+\beta _{\nu }^{2}}\right\}
d\lambda ;
\end{equation*}%
\begin{equation}
F_{n\mu N}^{2+}=\frac{J_{n}\left( \gamma _{jn}\right) }{f_{nN}^{s+}}\int
\left\{ \frac{\lambda ^{3}J_{n}^{\shortmid 2}\left( \lambda \right) }{\gamma
_{jn}^{2}-\lambda ^{2}}\left( 1-e^{-2\varkappa l}\right) \sum_{\nu
=N+1}^{\infty }\frac{\left( -1\right) ^{\nu }f_{n\nu }^{s+}}{\varkappa
^{2}+\beta _{\nu }^{2}}\right\} d\lambda ;  \label{55}
\end{equation}%
\begin{equation*}
G_{n\mu N}^{\ast 2+}=\frac{J_{n}\left( \gamma _{jn}\right) }{g_{nN}^{\ast s+}%
}\int \left\{ \frac{\lambda ^{3}J_{n}^{\shortmid 2}\left( \lambda \right) }{%
\varkappa \left( \gamma _{jn}^{2}-\lambda ^{2}\right) }\left(
1+e^{-2\varkappa l}\right) \sum_{j=N+1}^{\infty }\frac{g_{nj}^{\ast
s+}J_{n}\left( \gamma _{jn}\right) }{\gamma _{jn}^{2}-\lambda ^{2}}\right\}
d\lambda ;
\end{equation*}%
\begin{equation*}
G_{n\mu N}^{2+}=-\frac{J_{n}\left( \gamma _{jn}\right) }{g_{nN}^{s+}}\int
\left\{ \frac{\lambda ^{3}J_{n}^{\shortmid 2}\left( \lambda \right) }{\left(
\gamma _{jn}^{2}-\lambda ^{2}\right) }\left( 1-e^{-2\varkappa l}\right)
\sum_{j=N+1}^{\infty }\frac{g_{nj}^{s+}J_{n}\left( \gamma _{jn}\right) }{%
\gamma _{jn}^{2}-\lambda ^{2}}\right\} d\lambda ;
\end{equation*}%
\begin{equation*}
F_{n\mu N}^{\ast 1-}=\frac{\left( -1\right) ^{\mu }\beta _{\mu }}{%
f_{nN}^{\ast s-}}\int \left\{ \frac{\lambda J_{n}^{2}\left( \lambda \right)
}{\varkappa \left( \varkappa ^{2}+\beta _{\mu }^{2}\right) }\left(
1-e^{-2\varkappa l}\right) \sum_{\nu =N+1}^{\infty }\frac{\left( -1\right)
^{\nu }\beta _{\nu }f_{n\nu }^{\ast s-}}{\varkappa ^{2}+\beta _{\nu }^{2}}%
\right\} d\lambda ;
\end{equation*}%
\begin{equation*}
F_{n\mu N}^{1-}=\frac{\left( -1\right) ^{\mu }\beta _{\mu }}{f_{nN}^{s-}}%
\int \left\{ \frac{\lambda J_{n}^{\shortmid }\left( \lambda \right)
J_{n}\left( \lambda \right) }{\varkappa \left( \varkappa ^{2}+\beta _{\mu
}^{2}\right) }\left( 1-e^{-2\varkappa l}\right) \sum_{\nu =N+1}^{\infty }%
\frac{\left( -1\right) ^{\nu }\beta _{\nu }f_{n\nu }^{\ast s-}}{\varkappa
^{2}+\beta _{\nu }^{2}}\right\} d\lambda ;
\end{equation*}%
\begin{equation*}
G_{n\mu N}^{\ast 1-}=\frac{\left( -1\right) ^{\mu }\beta _{\mu }}{%
g_{nN}^{\ast s-}}\int \left\{ \frac{\lambda ^{2}J_{n}^{\shortmid }\left(
\lambda \right) J_{n}\left( \lambda \right) }{\varkappa \left( \varkappa
^{2}+\beta _{\mu }^{2}\right) }\left( 1-e^{-2\varkappa l}\right)
\sum_{j=N+1}^{\infty }\frac{g_{nj}^{\ast s-}J_{n}\left( \gamma _{jn}\right)
}{\gamma _{jn}^{2}-\lambda ^{2}}\right\} d\lambda ;
\end{equation*}%
\begin{equation*}
G_{n\mu N}^{1-}=-\frac{\left( -1\right) ^{\mu }\beta _{\mu }}{g_{nN}^{s-}}%
\int \left\{ \frac{\lambda ^{2}J_{n}^{\shortmid }\left( \lambda \right)
J_{n}\left( \lambda \right) }{\left( \varkappa ^{2}+\beta _{\mu }^{2}\right)
}\left( 1-e^{-2\varkappa l}\right) \sum_{j=N+1}^{\infty }\frac{%
g_{nj}^{s-}J_{n}\left( \gamma _{jn}\right) }{\gamma _{jn}^{2}-\lambda ^{2}}%
\right\} d\lambda ;
\end{equation*}%
\begin{equation*}
F_{n\mu N}^{\ast 2-}=\frac{J_{n}\left( \gamma _{jn}\right) }{f_{nN}^{\ast s-}%
}\int \left\{ \frac{\lambda ^{2}J_{n}^{\shortmid }\left( \lambda \right)
J_{n}\left( \lambda \right) }{\varkappa \left( \gamma _{jn}^{2}-\lambda
^{2}\right) }\left( 1-e^{-2\varkappa l}\right) \sum_{\nu =N+1}^{\infty }%
\frac{\left( -1\right) ^{\nu }\beta _{\nu }f_{n\nu }^{\ast s-}}{\varkappa
^{2}+\beta _{\nu }^{2}}\right\} d\lambda ;
\end{equation*}%
\begin{equation*}
F_{n\mu N}^{2-}=\frac{J_{n}\left( \gamma _{jn}\right) }{f_{nN}^{s-}}\int
\left\{ \frac{\lambda ^{3}J_{n}^{\shortmid 2}\left( \lambda \right) }{%
\varkappa \left( \gamma _{jn}^{2}-\lambda ^{2}\right) }\left(
1-e^{-2\varkappa l}\right) \sum_{\nu =N+1}^{\infty }\frac{\left( -1\right)
^{\nu }\beta _{\nu }f_{n\nu }^{s-}}{\varkappa ^{2}+\beta _{\nu }^{2}}%
\right\} d\lambda ;
\end{equation*}%
\begin{equation*}
G_{n\mu N}^{\ast 2-}=\frac{J_{n}\left( \gamma _{jn}\right) }{g_{nN}^{\ast s-}%
}\int \left\{ \frac{\lambda ^{3}J_{n}^{\shortmid 2}\left( \lambda \right) }{%
\varkappa \left( \gamma _{jn}^{2}-\lambda ^{2}\right) }\left(
1-e^{-2\varkappa l}\right) \sum_{j=N+1}^{\infty }\frac{g_{nj}^{\ast
s-}J_{n}\left( \gamma _{jn}\right) }{\gamma _{jn}^{2}-\lambda ^{2}}\right\}
d\lambda ;
\end{equation*}%
\begin{equation*}
G_{n\mu N}^{\ast 2-}=-\frac{J_{n}\left( \gamma _{jn}\right) }{g_{nN}^{s-}}%
\int \left\{ \frac{\lambda ^{3}J_{n}^{\shortmid 2}\left( \lambda \right) }{%
\left( \gamma _{jn}^{2}-\lambda ^{2}\right) }\left( 1+e^{-2\varkappa
l}\right) \sum_{j=N+1}^{\infty }\frac{g_{nj}^{s-}J_{n}\left( \gamma
_{jn}\right) }{\gamma _{jn}^{2}-\lambda ^{2}}\right\} d\lambda .
\end{equation*}%
where
\begin{equation*}
\frac{2^{1/6}l^{1/6}\text{$\Gamma $}\left( 2/3\right) \text{$\Gamma $}\left(
1/2\right) I_{1/6}\left( \varkappa l\right) }{\varkappa ^{7/6}sh\left(
\varkappa l\right) }-\sum_{\nu =0}^{N-1}\frac{f_{n\nu }^{\ast s+}\left(
-1\right) ^{\nu }}{\varkappa ^{2}+\beta _{\nu }^{2}}=\sum_{\nu =N+1}^{\infty
}\frac{f_{n\nu }^{\ast s+}\left( -1\right) ^{\nu }}{\varkappa ^{2}+\beta
_{\nu }^{2}};
\end{equation*}%
\begin{equation*}
\frac{2^{7/6}l^{7/6}\text{$\Gamma $}\left( 5/3\right) \text{$\Gamma $}\left(
1/2\right) I_{7/6}\left( \varkappa l\right) }{\varkappa ^{13/6}sh\left(
\varkappa l\right) }-\sum_{\nu =0}^{N-1}\frac{f_{n\nu }^{s+}\left( -1\right)
^{\nu }}{\varkappa ^{2}+\beta _{\nu }^{2}}=\sum_{\nu =N+1}^{\infty }\frac{%
f_{n\nu }^{s+}\left( -1\right) ^{\nu }}{\varkappa ^{2}+\beta _{\nu }^{2}};
\end{equation*}%
\begin{equation*}
\frac{2^{2/3}\lambda ^{1/3}\text{$\Gamma $}\left( 2/3\right) J_{n+5/3}\left(
\lambda \right) }{\left( \lambda ^{2}-n^{2}\right) J_{n}^{2}\left( \lambda
\right) J_{n}^{\shortmid }\left( \lambda \right) }-\sum_{\nu =0}^{N-1}\frac{%
g_{n\nu }^{\ast s\pm }J_{n}\left( \gamma _{jn}\right) }{\gamma
_{jn}^{2}-\lambda ^{2}}=\sum_{\nu =N+1}^{\infty }\frac{g_{n\nu }^{\ast s\pm
}J_{n}\left( \gamma _{jn}\right) }{\gamma _{jn}^{2}-\lambda ^{2}};
\end{equation*}%
\begin{equation*}
\frac{2^{2/3}\lambda ^{-2/3}\text{$\Gamma $}\left( 5/3\right) \text{$\Gamma $%
}\left( 1/2\right) J_{n+2/3}\left( \lambda \right) }{\left( \lambda
^{2}-n^{2}\right) J_{n}^{2}\left( \lambda \right) J_{n}^{\shortmid }\left(
\lambda \right) }-\sum_{\nu =0}^{N-1}\frac{g_{n\nu }^{s\pm }J_{n}\left(
\gamma _{jn}\right) }{\gamma _{jn}^{2}-\lambda ^{2}}=\sum_{\nu =N+1}^{\infty
}\frac{g_{n\nu }^{s\pm }J_{n}\left( \gamma _{jn}\right) }{\gamma
_{jn}^{2}-\lambda ^{2}};
\end{equation*}%
\begin{equation*}
\frac{2^{1/3}\text{$\Gamma $}\left( 2/3\right) \Gamma \left( 1/2\right) %
\left[ \varkappa lI_{1/6}^{\shortmid }\left( \varkappa l\right)
-1/6I_{1/6}\left( \varkappa l\right) \right] }{\varkappa ^{1/6}sh\left(
\varkappa l\right) }-\sum_{\nu =0}^{N-1}\frac{f_{n\nu }^{\ast s-}\left(
-1\right) ^{\nu }\beta _{\nu }}{\varkappa ^{2}+\beta _{\nu }^{2}}=\sum_{\nu
=N+1}^{\infty }\frac{f_{n\nu }^{\ast s-}\left( -1\right) ^{\nu }\beta _{\nu }%
}{\varkappa ^{2}+\beta _{\nu }^{2}}
\end{equation*}%
\begin{equation*}
\frac{2^{7/6}l^{13/6}\text{$\Gamma $}\left( 5/3\right) \text{$\Gamma $}%
\left( 1/2\right) \left[ \varkappa lI_{7/6}^{\shortmid }\left( \varkappa
l\right) -7/6I_{7/6}\left( \varkappa l\right) \right] }{\varkappa
^{13/6}sh\left( \varkappa l\right) }-\sum_{\nu =0}^{N-1}\frac{f_{n\nu
}^{s+}\left( -1\right) ^{\nu }\beta _{\nu }}{\varkappa ^{2}+\beta _{\nu }^{2}%
}=\sum_{\nu =N+1}^{\infty }\frac{f_{n\nu }^{s+}\left( -1\right) ^{\nu }\beta
_{\nu }}{\varkappa ^{2}+\beta _{\nu }^{2}}
\end{equation*}%
Substituting the obtained values into condition of hydro-resilient contact (%
\ref{9a}) and taking into account (\ref{23}), (\ref{33}), (\ref{34}), (\ref%
{55}) we obtain such unlimited systems written in matrix form to evaluate
the unknown coefficients
\begin{equation*}
f_{n\nu }^{e\pm },f_{n\nu }^{\ast e\pm },g_{nj}^{e\pm },g_{nj}^{\ast e\pm
},f_{n\nu }^{o\pm },f_{n\nu }^{\ast o\pm },g_{nj}^{o\pm },g_{nj}^{\ast o\pm
};\left( \nu ,j=0,1,2......N\right) ;
\end{equation*}%
\begin{equation}
\underset{N\times N}{\left( F_{1n}^{1\pm }\right) _{\nu \mu }}\underset{%
1\times N}{\left\{ f_{n}^{e\pm }\right\} _{\nu }}+\underset{N\times N}{%
\left( F_{2n}^{\ast 1\pm }\right) _{\nu \mu }}\underset{1\times N}{\left\{
f_{n}^{\ast e\pm }\right\} _{\nu }}+\underset{N\times N}{\left( G_{1n}^{1\pm
}\right) _{\nu \mu }}\underset{1\times N}{\left\{ g_{n}^{e\pm }\right\} _{j}}%
+\underset{N\times N}{\left( G_{2n}^{\ast 1\pm }\right) _{\nu \mu }}\underset%
{1\times N}{\left\{ g_{n}^{\ast e\pm }\right\} _{j}}=0;  \label{56}
\end{equation}%
\begin{equation}
\underset{N\times N}{\left( F_{1n}^{2\pm }\right) _{\nu \mu }}\underset{%
1\times N}{\left\{ f_{n}^{e\pm }\right\} _{\nu }}+\underset{N\times N}{%
\left( F_{2n}^{\ast 2\pm }\right) _{\nu \mu }}\underset{1\times N}{\left\{
f_{n}^{\ast e\pm }\right\} _{\nu }}+\underset{N\times N}{\left( G_{1n}^{2\pm
}\right) _{\nu \mu }}\underset{1\times N}{\left\{ g_{n}^{e\pm }\right\} _{j}}%
+\underset{N\times N}{\left( G_{2n}^{\ast 2\pm }\right) _{\nu \mu }}\underset%
{1\times N}{\left\{ g_{n}^{\ast e\pm }\right\} _{j}}=0  \label{57}
\end{equation}%
\begin{equation}
\underset{N\times N}{\left( R_{1n}^{\pm }\right) _{\nu \mu }}\underset{%
1\times N}{\left\{ f_{rn}^{\pm }\right\} _{\nu }}+\underset{}{\underset{%
N\times N}{\left( R_{2n}^{\pm }\right) _{\nu \mu }}\underset{1\times N}{%
\left\{ g_{zn}^{\pm }\right\} _{j}}}=\underset{1\times N}{\left\{
f_{n}^{\ast e\pm }\right\} _{\nu }}+\underset{1\times N}{\left\{ f_{n}^{\ast
a\pm }\right\} _{\nu }};  \label{58}
\end{equation}%
\begin{equation}
\underset{N\times N}{\left( R_{3n}^{\pm }\right) _{\nu \mu }}\underset{%
1\times N}{\left\{ f_{rn}^{\pm }\right\} _{\nu }}+\underset{}{\underset{%
N\times N}{\left( R_{4n}^{\pm }\right) _{\nu \mu }}\underset{1\times N}{%
\left\{ g_{zn}^{\pm }\right\} _{j}}}=\underset{1\times N}{\left\{
g_{n}^{\ast e\pm }\right\} _{j}}+\underset{1\times N}{\left\{ g_{n}^{\ast
a\pm }\right\} _{j}};  \label{59}
\end{equation}%
\begin{equation}
\underset{N\times N}{\left( R_{1n}^{\pm }\right) _{\nu \mu }}\underset{%
1\times N}{\left\{ f_{rn}^{\pm }\right\} _{\nu }}+\underset{}{\underset{%
N\times N}{\left( R_{2n}^{\pm }\right) _{\nu \mu }}\underset{1\times N}{%
\left\{ g_{zn}^{\pm }\right\} _{j}}}=\underset{N\times N}{\left( E_{1n}^{\pm
}\right) _{\nu \mu }}\underset{1\times N}{\left\{ f_{n}^{\ast o\pm }\right\}
_{\nu }};  \label{60}
\end{equation}%
\begin{equation}
\underset{N\times N}{\left( R_{3n}^{\pm }\right) _{\nu \mu }}\underset{%
1\times N}{\left\{ f_{rn}^{\pm }\right\} _{\nu }}+\underset{}{\underset{%
N\times N}{\left( R_{4n}^{\pm }\right) _{\nu \mu }}\underset{1\times N}{%
\left\{ g_{zn}^{\pm }\right\} _{j}}}=\underset{N\times N}{\left( E_{2n}^{\pm
}\right) _{\nu \mu }}\underset{1\times N}{\left\{ f_{n}^{\ast o\pm }\right\}
_{\nu }}  \label{61}
\end{equation}%
\begin{equation}
\underset{N\times N}{\left( E_{3n}^{\pm }\right) _{\nu \mu }}\underset{%
1\times N}{\left\{ f_{n}^{\ast o\pm }\right\} _{\nu }}+\underset{N\times N}{%
\left( M_{1n}^{\pm }\right) _{j\mu }}\underset{1\times N}{\left\{
g_{n}^{\ast o\pm }\right\} _{j}}=\underset{1\times N}{\left\{ f_{n}^{o\pm
}\right\} _{\nu }};  \label{62}
\end{equation}%
\begin{equation}
\underset{N\times N}{\left( E_{4n}^{\pm }\right) _{\nu \mu }}\underset{%
1\times N}{\left\{ f_{n}^{\ast o\pm }\right\} _{\nu }}+\underset{N\times N}{%
\left( M_{2n}^{\pm }\right) _{j\mu }}\underset{1\times N}{\left\{
g_{n}^{\ast o\pm }\right\} _{j}}=\underset{1\times N}{\left\{ g_{n}^{o\pm
}\right\} _{\nu }}  \label{63}
\end{equation}%
where
\begin{equation*}
\overline{R}_{in\nu \mu }=\rho \omega ^{2}c^{2}R_{in\nu \mu };i=1,2,3,4.
\end{equation*}%
\begin{equation*}
f_{rn\nu }^{\pm }=f_{n\nu }^{e\pm }+f_{n\nu }^{a\pm }-f_{n\nu }^{o\pm
};g_{znj}^{\pm }=g_{nj}^{e\pm }+g_{nj}^{a\pm }-g_{nj}^{o\pm };
\end{equation*}%
\begin{equation*}
E_{4n\mu \nu }^{-}=\left[ \sum_{\nu =1}^{N}C_{n\nu \mu }^{-}J_{n}\left(
\gamma _{jn}\right) \left( -1\right) ^{\nu }\beta _{\nu }+\frac{2N}{l}\right]
\delta _{\nu \mu };
\end{equation*}%
\begin{equation*}
E_{3n\mu \nu }^{\pm }=J_{n}\left( \varkappa _{2}\right) \delta _{\nu \mu
};E_{2n\mu \nu }^{\pm }=\frac{\beta _{0}}{\xi _{02}\varepsilon _{n\mu }}%
\delta _{\nu \mu };E_{1n\mu \nu }^{\pm }=\frac{\varkappa
_{2}J_{n}^{\shortmid }\left( \varkappa _{2}\right) }{\xi _{01}\beta _{0}^{-1}%
};
\end{equation*}%
\begin{equation*}
E_{4n\mu \nu }^{+}=\left\{ \left[ \sum_{\nu =1}^{N}C_{n\nu \mu
}^{+}\varepsilon _{\mu n}\left( -1\right) ^{\nu }\right] +\frac{2}{\pi l}%
\sum_{\nu =N+1}^{\infty }\frac{1}{\nu ^{2}}\right\} \delta _{\nu \mu }
\end{equation*}%
\begin{equation*}
M_{1n\mu \nu }^{\pm }=C_{n\nu \mu }^{\pm }J_{n}\left( \gamma _{jn}\right)
;M_{2n\mu \nu }^{+}=\frac{\left( -1\right) ^{\nu }\varkappa
_{2}J_{n}^{\shortmid }\left( \varkappa _{2}\right) J_{n}\left( \gamma
_{jn}\right) \varepsilon _{\mu n}}{\gamma _{\mu n}^{2}-\varkappa _{2}^{2}};
\end{equation*}%
\begin{equation*}
M_{2n\mu \nu }^{+}=\frac{\left( -1\right) ^{\nu }\varkappa
_{2}J_{n}^{\shortmid }\left( \varkappa _{2}\right) J_{n}\left( \gamma
_{jn}\right) \varepsilon _{\mu n}\beta _{\nu }}{\gamma _{\mu
n}^{2}-\varkappa _{2}^{2}}.
\end{equation*}

Thus we obtain the closed finite system of $8\left( 2N+1\right) $ linear
algebraic equations for the symmetrical and anti-symmetrical components of
the solution of given problem. From the system (\ref{56}) -(\ref{63}) we
define the unknown coefficients
\begin{equation*}
f_{n\nu }^{e\pm },f_{n\nu }^{\ast e\pm },g_{nj}^{e\pm },g_{nj}^{\ast e\pm
},f_{n\nu }^{o\pm },f_{n\nu }^{\ast o\pm },g_{nj}^{o\pm },g_{nj}^{\ast o\pm
};\nu ,j=0,1,2....N
\end{equation*}%
Let us consider some particular cases, which are directly obtained from (\ref%
{56}):

1) If liquid filler is absent inside the cylindrical cover than we obtain
the soluted system of first four matrix equations (\ref{56}) -(\ref{59})
with the following modifications
\begin{equation*}
f_{n\nu }^{o\pm }=0;g_{nj}^{o\pm }=0;
\end{equation*}

2) For absolute solid (soft) cylinder the soluted system will be formed of
matrix equations (\ref{56}), (\ref{57}) and such two equations
\begin{equation}
\underset{N\times 1}{\left\{ f_{n}^{\ast e\pm }\right\} _{\nu }=}\underset{%
N\times 1}{\left\{ f_{n}^{\ast a\pm }\right\} _{\nu };}\underset{N\times 1}{%
\left\{ g_{n}^{\ast e\pm }\right\} _{j}=}\underset{N\times 1}{\left\{
g_{n}^{\ast a\pm }\right\} _{j};}\underset{N\times 1}{\left\{ f_{n}^{e\pm
}\right\} _{\nu }=}\underset{N\times 1}{\left\{ f_{n}^{a\pm }\right\} _{\nu
};}\underset{N\times 1}{\left\{ g_{n}^{e\pm }\right\} _{j}=}\underset{%
N\times 1}{\left\{ g_{n}^{a\pm }\right\} _{j};}  \label{64}
\end{equation}

3) If the cylindrical cover is replaced with liquid cylinder we obtain the
soluted system containing (\ref{56}), (\ref{57}), (\ref{62}), (\ref{63})
matrix equations and such matrix equations:
\begin{equation}
\underset{N\times 1}{\left\{ f_{n}^{\ast e\pm }\right\} _{\mu }+}\underset{%
N\times 1}{\left\{ f_{n}^{\ast a\pm }\right\} _{\mu }=}\underset{}{\underset{%
N\times N}{\left( E_{1N}^{\pm }\right) _{\nu \mu }}\underset{N\times 1}{%
\left\{ f_{n}^{\ast o\pm }\right\} }_{\mu \nu };}\underset{N\times 1}{%
\left\{ g_{n}^{\ast e\pm }\right\} _{\mu }+}\underset{N\times 1}{\left\{
g_{n}^{\ast a\pm }\right\} _{\mu }=}\underset{}{\underset{N\times N}{\left(
E_{2N}^{\pm }\right) _{j\mu }}\underset{N\times 1}{\left\{ g_{n}^{\ast o\pm
}\right\} _{\mu }};}  \label{65}
\end{equation}%
\begin{equation*}
\underset{N\times 1}{\left\{ f_{n}^{e\pm }\right\} _{\mu }+}\underset{%
N\times 1}{\left\{ f_{n}^{a\pm }\right\} _{\mu }}\underset{}{=}\underset{%
N\times 1}{\left\{ f_{n}^{o\pm }\right\} _{\mu };}\underset{N\times 1}{%
\left\{ g_{n}^{e\pm }\right\} _{\mu }+}\underset{N\times 1}{\left\{
g_{n}^{a\pm }\right\} _{\mu }}\underset{}{=}\underset{N\times 1}{\left\{
g_{n}^{o\pm }\right\} _{\mu };}
\end{equation*}

4) If point pressure source
\begin{equation*}
P^{o}=P_{0}\delta \left( r-r_{0}\right) \delta \left( \theta \right) /4\pi r;
\end{equation*}%
is placed inside the cylindrical cover than the \ system of matrix equations
(\ref{56}) - (\ref{63}) must be changed as follows
\begin{equation}
f_{n\nu }^{\ast a+}=f_{n\nu }^{\ast a-}=g_{nj}^{a-}=g_{nj}^{\ast a+}=0;
\label{66}
\end{equation}%
\begin{equation*}
g_{nj}^{a+}=-\sum_{\nu =0}^{\infty }\frac{P_{0}\varepsilon _{\nu
}\varepsilon _{nj}\left( -1\right) ^{\nu }J_{n}\left( \gamma _{jn}\right)
\cos \beta _{\nu }z_{0}}{\pi l};g_{nj}^{\ast a-}=-\sum_{\nu =0}^{\infty }%
\frac{P_{0}\beta _{\nu }\varepsilon _{nj}\left( -1\right) ^{\nu }J_{n}\left(
\gamma _{jn}\right) \sin \beta _{\nu }z_{0}}{\pi l};
\end{equation*}%
\begin{equation*}
f_{nj}^{a-}=-\sum_{j=0}^{\infty }\frac{P_{0}\varepsilon _{nj}J_{n}^{2}\left(
\gamma _{jn}\right) \sin \beta _{\nu }z_{0}}{\pi l};f_{nj}^{a+}=-\sum_{\nu
=0}^{\infty }\frac{P_{0}\varepsilon _{\nu }\varepsilon _{nj}J_{n}\left(
\gamma _{jn}\right) \cos \beta _{\nu }z_{0}}{\pi l}.
\end{equation*}

Thus, in this work a new problem solution methodology of contact interaction
of acoustic medium with resilient finite bodies of cylindrical form, based
on application of boundary integral equations method in conjunction with
series method with later use of series convergence improvement taking into
account the particularities of the determined functions.

\end{document}